\documentclass [11pt,letterpaper]{JHEP3}
\usepackage{amsmath}
\usepackage{epsfig}
\usepackage{epstopdf}
\usepackage{psfrag}
\usepackage{cite}       
\usepackage{bm}         
\usepackage{verbatim}   
\advance\parskip 1.0pt plus 1.0pt minus 2.0pt   
\def\href#1#2{#2}	

\def\coeff#1#2{{\textstyle {\frac {#1}{#2}}}}
\def\half{\coeff 12}

\def\N{{\cal N}}

\def\G{{\cal G}}
\def\S_1{{\widetilde {S_1}}}

\def\R{{\mathbb R}}

\def\tr{{\rm tr}}

\def\Z{{\mathbb Z}}

\def\Dslash{{\rlap{\raise 1pt \hbox{$\>/$}}D}}

\preprint{SLAC-PUB-12825}
\title
    {%
    \boldmath   Magnetic bion condensation: A new mechanism of confinement and mass gap
      in four dimensions  }
\author
    {%
    Mithat  \"Unsal$^1$\footnote{\email{unsal@slac.stanford.edu}}~
    \\${}^1$ SLAC, Stanford University, Menlo Park, California 94025
    \\\; Physics Department, Stanford University, Stanford, California  94305, USA
    }%
\abstract
    {%
In recent work, we derived  the long-distance confining dynamics of certain QCD-like gauge 
theories formulated  on small $S^1 \times \R^3$ based     on symmetries, an index theorem, 
and Abelian duality.    Here, we give the microscopic derivation. The solution reveals  a new mechanism of confinement in QCD(adj) in the regime where we have control over both perturbative and nonperturbative aspects. In particular, consider $SU(2)$ QCD(adj) theory with 
$1 \leq n_f \leq 4$ Majorana fermions, a theory which undergoes gauge  symmetry breaking at small $S^1$.   If the  magnetic charge  of the BPS  monopole is normalized to unity, we show that confinement occurs due to  condensation of objects with magnetic charge $2$, not $1$. Because of  index theorems,   we know that such an object cannot be a two identical monopole configuration. Its   net topological charge must vanish, and hence it must be topologically indistinguishable from the perturbative vacuum.   We construct such non-self-dual topological excitations, the magnetically charged, topologically null  molecules of a BPS monopole and  
${\overline{\rm KK}}$ antimonopole, which we   refer to  as magnetic bions.   An immediate puzzle with this proposal is the apparent Coulomb repulsion between the BPS-${\overline{\rm KK}}$ pair. An attraction which overcomes the Coulomb repulsion between the two   is induced  by $2n_f$-fermion  exchange. Bion condensation is also the mechanism of confinement in  $\N=1$ SYM on the same four-manifold.    The $SU(N)$ generalization  hints  a  possible 
 hidden integrability behind nonsupersymmetric  QCD of affine Toda type, and allows us to analytically compute the mass gap in the gauge sector.   
   We currently do not know the extension to $\R^4$.
          }%
\keywords{Nonperturbative QCD, Duality}
\begin{document}
\section{Introduction}
\label{sec:int}
Probably the most  
important experimental and phenomenological observation about $SU(3)$ 
QCD is confinement, i.e., the absence of the free colored particles in isolation.  Numerical lattice simulations unambiguously establish confinement in pure Yang-Mills theory and QCD.  However, 
to date the analytical success had been limited. For reviews, see \cite{Greensite:2003bk, 
Shifman:2007ce, Bruckmann:2007ru}

The QCD of nature belongs to a subclass of   asymptotically free and  confining  gauge theories   without  elementary scalars. This class is referred to as vectorlike or QCD-like.  This is a sufficiently good  reason to warrant the 
study of  the dynamics of  such four-dimensional QCD-like theories. 
In the last two decades, most theoretical  efforts were concentrated  into the dynamics of supersymmetric theories.  
  It would be fair to say that despite many remarkable results obtained in such theories, 
  their benefit to  QCD-like theories is still 
  in its infancy. 
   There is a very good reason for this.  On $\R^4$, there only exists one QCD-like 
  supersymmetric theory, the pure $\N=1$ SYM.  All other supersymmetric theories have scalars, and are hence non-QCD-like by definition.  In regimes where such theories are solved or understood quantitatively,   such as mass 
  deformation of $\N=2$ SYM down to $\N=1$ \cite{Seiberg:1994rs}, the  scalars 
  never decouple from the dynamics. If they are forced  to decouple  by tuning certain parameters, one usually looses  the theoretical control over the theory \cite{Douglas:1995nw}.  
  
Our goal in  this paper is more direct and motivated by the following question:

\begin{quote} {\it
 Is there any asymptotically free and confining 
QCD-like theory in $d=4$ dimensions (with no special properties such as supersymmetry) in which we 
can understand its   nonperturbative aspects   exactly, and  can  derive the long-distance (confining) dynamics starting with microscopic theory?} \footnote{Two archetypes  of 
non-QCD-like  theory in which the long-distance theory can be derived starting with the microscopic theory are 
Polyakov's treatment    \cite{Polyakov:1976fu} of  the Georgi-Glashow model on  $\R^3$, a theory which confines,   and   Nekrasov's derivation  
 \cite{Nekrasov:2002qd} of the $\N=2$  Seiberg-Witten prepotential, a theory which does not  confine.  The  $\N=1$ mass deformation of the latter also confines.  
 Our goal is to find such quantitatively tractable   examples among QCD-like theories. }
 \end{quote}

On $\R^4$,  the answer seems to be out of reach currently.     However, on locally 
four-dimensional 
settings, such as spatial $S^1 \times \R^3 $,  the answer is yes. In particular,  
 QCD with multiple adjoint representation fermions on small 
$S^1 \times \R^3$,  ($S^1 \times \R^{2,1}$ in Minkowski setting) \cite{Unsal:2007vu}  becomes 
analytically tractable.  
 Here, it is important that  $S^1$ is not a thermal circle.  It is   a spatial circle  
 along which fermions are endowed with periodic spin connection,  and the resulting QCD-like theory  
 is a zero temperature field theory on a space with one compact dimension.   
The benefits of   considering this setup are  
 $ {(\it i)}$  weak coupling (due to asymptotic freedom) and    
 ${(\it  ii)}$ unbroken spatial center symmetry.   
The latter is a consequence of the absence of thermal fluctuations and the fact that the quantum fluctuations favor the center symmetric vacuum.

\subsection{A mechanism of confinement by non-self-dual topological excitations}
 Below, we will briefly outline the results of \cite{Unsal:2007vu}, and  address fundamental   questions regarding the microscopic origin of confinement and chiral symmetry realization in QCD(adj).   Historically, (starting with the mid-1970s), confinement was thought to be related to {\bf self-dual} topological 
excitations which are solutions to Prasad-Sommerfield--type   first-order differential equations. 
For example, both the Polyakov model \cite{Polyakov:1976fu} and Seiberg-Witten theory are of this type \cite{Seiberg:1994rs}. In particular,  the Seiberg-Witten solution may be viewed as a very elegant 
realization of the mid-1970's dream of 't Hooft, Polyakov and Mandelstam. The picture of confinement (which appears in a semiclassical regime) of QCD(adj) does not directly fit to  earlier ideas regarding the subject. As we will demonstrate  below, it is sourced by {\bf non-self-dual}, yet dynamically stable novel topological excitations, that we will refer to as {\bf magnetic bions}. What makes these excitations  more  elusive than monopole-instantons, monopoles or instantons is that, magnetic bions, in the sense of topological charge, are indistinguishable from perturbative vacuum. Thus, there was no reason  to search for their existence.  This is also what makes the current work different from earlier  (above-mentioned) proposals of confinement in non-abelian gauge theories.  

  Consider the setup of Ref.\cite{Unsal:2007vu}. 
In the small $S^1$ (weak coupling)  limit of  $SU(2)$ QCD(adj),    
the holonomy of the spatial Wilson line along the $S^1$ direction $U({ x}) = P e^{i\int dx_4 A_4({x}, x_4)}$   may be regarded  as a {\it compact} 
adjoint Higgs field.  This field acquires a nontrivial (center symmetry respecting) 
vacuum expectation value, $ U= {\rm Diag}( e^{i \pi/2}, e^{-i \pi/2} )$, due to 
radiatively induced one-loop Coleman-Weinberg  potential.    
The photons and neutral 
 fermions   $(A_{\mu},  \lambda^{I})$
 parallel to $U$  remains massless to {\it all} orders in perturbation theory, and all the other modes acquire masses and hence decouple from the infrared dynamics.  
  
Nonperturbatively,   there are topologically stable monopole configurations which are  a consequence of gauge symmetry breaking. 
Since the adjoint Higgs field is compact,  other than the Bogomol'nyi-Prasad-Sommerfield 
(BPS) monopole, there is also a KK monopole. The existence of KK-monopoles, which are perhaps the most  crucial ingredient in our discussion of QCD(adj), was discovered  in 1997, independently  by Lee and Yi using D-branes in string theory  \cite{Lee:1997vp} and  by Kraan and van Baal by using calorons configurations   \cite{Kraan:1998pm}.  
  The magnetic 
and topological charges $ \left( \int  F,  \; \int  F \tilde F \right) $ of these monopoles are normalized as 
\begin{equation}
{\rm BPS:} (+1,+ \half), \qquad {\overline {\rm BPS}}: (-1, -\half) \qquad {\rm KK:} (-1, +\half), \qquad 
{\overline {\rm KK}}: (+1, -\half) 
\end{equation}
where bar denotes antimonopoles. 

In  \cite{Unsal:2007vu}, we constructed the $d=3$ dimensional   long-distance theory for 
 QCD(adj) formulated on $\R^3 \times S^1$  by employing  three tools:   
 abelian duality,  symmetries, and index theorem. This strategy is, in essence, similar to the 
 Seiberg-Witten construction of prepotential in  $\N=2$ SYM  \cite{Seiberg:1994rs}. 
   The unique lagrangian   to order $e^{-2S_0}$   dictated by these considerations is 
\begin{eqnarray}
L^{\rm dQCD} = \frac{1}{2} (\partial \sigma)^2 -  b\;  e^{-2S_0}\cos 2 \sigma  
+  i \bar \psi^I \gamma_{\mu} \partial_{\mu} \psi_I   
 + c \; e^{-S_0}  \cos \sigma   
( \det_{I, J} \psi^I \psi^J +  \rm c.c.) \qquad 
\label{Eq:dQCD}
\end{eqnarray}
where $\sigma$ and $\psi^I$  denote    the dual photon and fermion. Dimensionless coordinates, measured in units of compactification circumference $L$,  are used.     A detailed microscopic 
derivation of this Lagrangian will be given in section \ref{sec:dyn}.
The mass gap for gauge bosons is manifest in this lagrangian. The inverse of the mass gap is the characteristic size of the 
chromoelectric flux tube, hence  confinement is also manifest in dual formulation \cite{Unsal:2007vu}. 
  
  \subsection{Microscopic derivation}
In this work, we will derive the dual  lagrangian \ref{Eq:dQCD} by summing over all non-perturbative effects. Before doing so,   
note a simple but important  feature of \ref{Eq:dQCD}.  It is clear 
that fermionic interaction terms arise due to  the monopole effects.  Any  monopole carries a net topological charge. If massless fermions are present in the underlying theory,  due to the index theorem,   
a monopole must be  associated with $2n_f$ fermion zero-modes of one chirality and an antimonopole leads to  $2n_f$  zero-modes of the opposite chirality.  
Consequently, the terms involving fermion zero-mode insertions are the sum of the monopole operators:  
\begin{eqnarray}
&& {\rm BPS}: e^{ i \sigma}  \det_{I,J} \psi^I \psi^J,  \qquad {\rm KK}: e^{ - i \sigma}  \det_{I,J} \psi^I \psi^J,  \cr 
&& {\rm \overline {BPS}}: e^{ -i \sigma}  \det_{I,J} \bar \psi^I \bar \psi^J,  \qquad {\rm \overline{KK}}:
  e^{ i \sigma}  \det_{I,J} \bar \psi^I \bar \psi^J,  
\end{eqnarray}
where  $e^{iq_m \sigma}$ is the pure monopole  operator and $q_m= \pm 1$ are magnetic charges of the corresponding  (anti)monopole, and  $\det_{I,J} \psi^I \psi^J$ are compulsory 
zero-modes attached to it.  
Now, let us inspect   the bosonic potential.  It is 
\begin{eqnarray}
V(\sigma) \sim  \cos 2 \sigma \sim e^{ i 2\sigma}  + e^{-2i \sigma}
\end{eqnarray}
Because of the index theorem,  a bosonic potential cannot arise due to objects which carry a nonvanishing index.
 Such objects, by construction, must have fermion zero-mode insertions,
  and cannot appear in the bosonic potential.
It is easy to check that the functional integral $Z= \int D\sigma \;  e^{- \int d^3x \; \left[ \frac{1}{2} (\partial \sigma)^2 -  b\;  e^{-2S_0}\cos 2 \sigma \right]}$  
 is equivalent to a  plasma  of magnetically charged particles with long range 
Coulomb interaction, 
\begin{equation}
V(r) = \frac{ 2(\pm 2) }{4 \pi r} 
\end{equation}
where charges are twice the one of the monopoles.
In other words,  the Debye phenomena (which renders  the dual photon massive) is induced not due to 
excitations  with magnetic and topological charge $(\pm 1, \pm \half) $, but rather with 
 charges $(\pm 2, 0)$.  Clearly, these are not elementary monopoles.  The first question we want to answer is,    what are these objects? 

A fuller discussion of all pairs and their roles will be given in section 2.2.  For now, 
let us observe that only  a bound state of  BPS monopole, and KK antimonopole,  
  BPS$\overline {\rm KK}$ , and its antiparticle can induce the bosonic potential.  
   Such an  object has the correct   quantum numbers $(1, \half) + (1, -\half)= (2, 0)$
   and is the prime candidate for the magnetically charged object which  leads to confinement in QCD(adj) in the $L\Lambda \ll 1$ regime.

There is an immediate  puzzle    with this proposal.  The BPS and $\overline {\rm KK}$  monopoles 
interact electromagnetically  via Coulomb repulsion,  hence in order to have a bound state, there  must exist an attraction which may overcome the Coulomb repulsion.\footnote{This  situation is analogous to the BCS theory of superconductivity.  There must exist a net attraction between electron pairs which overcomes the 
shielded, yet repulsive Coulomb potential. Such an attractive  force is provided through 
the exchange of phonons  of the crystal lattice.  A novel   pairing mechanism is at work in QCD(adj) formulated on small $S^1 \times \R^3$. As  will be seen explicitly, the pairing in QCD(adj) is a real-space phenomena, unlike the BCS theory.}  
In the QCD(adj) vacuum, a pairing mechanism   arguably  as strange as   the BCS theory 
\cite{Bardeen:1957mv}
 takes place .   An attraction  
which overwhelms the Coulomb repulsion  between BPS and $\overline {\rm KK}$
is generated via (an even number of) fermion exchange.  In $n_f=1$ QCD(adj) (i.e., SYM), this is a  fermion pair exchange.  In $n_f >1$ QCD(adj), it is the exchange of $2n_f$ fermions.   The attractive 
 potential is a  logarithmical  one 
 \begin{equation}
 V_{\rm eff} (r) = 4n_f \log r + \frac{ 1}{4 \pi r}, \qquad  r \gg 1 
\label{Eq:effective}
 \end{equation}
and it easily  overcomes the repulsive Coulomb force. 
 This forces the BPS and $\overline {\rm KK}$  monopoles to form a charged bound state.  
We refer to this molecule as a {\bf  magnetic bion}, and to the $\overline {\rm BPS}$-KK molecule as 
{\bf anti-bion}.   The important point that is worth repeating is that 
the net topological charge of the  BPS-$\overline {\rm KK}$ pair  is identically zero: 
$
\int_{\R^3 \times S^1}  F \widetilde F = 0
$, 
even though for individual (isolated) BPS  it is  $\int_{\R^3 \times S^1} F \widetilde F = \frac{1}{2}$,
and for $\overline {\rm KK}$ it is   $\int_{\R^3 \times S^1}  F \widetilde F = - \frac{1}{2}$. Consequently, 
bions do not have fermions zero-modes attached to  them, and 
they  are the leading contribution  to the effective bosonic potential for the dual photon.

Considerations along these line also provides dynamical explanations for the absence of confinement 
in Yang-Mills {\it noncompact} Higgs system with adjoint Dirac fermions 
formulated on  $\R^3$.  
Affleck, Harvey and Witten  in Ref.\cite{Affleck:1982as} showed that such systems do not confine despite the presence of magnetic monopoles. Their argument is based on symmetries and index 
theorems.  Without much recourse to  the microscopic theory, 
  they showed that 
the photon arises as a Goldstone boson of spontaneously broken fermion number symmetry, 
hence  remains massless nonperturbatively.    Here, we give  a microscopic derivation of this beautiful symmetry argument based on the dynamics of monopoles (and bions).  In one sentence, 
the absence of magnetically charged, but topologically null configurations (which may be the only source of a mass gap  for a dual photon in the presence of fermions) implies the absence of confinement 
in the $SU(2)$ application.  We also provide  a dynamical explanation for the absence of confinement in 
the $\N=2$ SYM theory on $\R^3$ based on a  similar rationale \footnote{These theories 
(formulated on $\R^3$)  are as important as QCD(adj) on 
$\R^3 \times S^1$. They exhibit that  if massless fermions are present,  
having monopoles is not sufficient  to have confinement. }.

The discussion of  nonsupersymmetric QCD(adj) can also be applied to $\N=1$ SYM on 
$\R^3 \times S^1$  with only  cosmetic changes. All one needs to be careful about is the extra massless scalar, and keep it in the effective theory.  In fact, the long-distance effective theory for SYM (which is a supersymmetric affine Toda theory)  was derived far before  our work on the subject 
\cite{Katz:1996th, Seiberg:1996nz, Davies:1999uw, Davies:2000nw}. \footnote{ Our derivation of the bosonic potentials in SYM differs  from earlier work, which was based on using supersymmetry as a completion device to obtain superpotential (hence bosonic potential) from 
the monopole induced fermionic terms.  We instead chose to delineate  on the microscopic origin 
of the bosonic potential, and obtained it directly without any recourse to supersymmetry. 
 The final result is  the same  of  earlier work \cite{Katz:1996th, Seiberg:1996nz, Davies:1999uw, Davies:2000nw}.  The real payoff of our  approach is in  its applicability to  nonsupersymmetric theories.} 
In spite of that, the fact that confinement 
was induced not due to (self-dual)  monopoles, but rather via  (non-self-dual)
magnetic bions  was not understood earlier.   Remarkably, the mechanism of confinement 
for $\N=1$ SYM and nonsupersymmetric QCD(adj) is one and the same in the small $S^1$ regime.

 The second part of the paper  discusses  the $SU(N)$  generalization of the 
 nonsupersymmetric QCD(adj), and derives the long-distance Lagrangian.  The biggest surprise is that the   bosonic sector of QCD(adj) maps into  an integrable system, 
 intimately related to possible integrable   generalization of the    affine Toda theories.    We identify magnetic bions as bound states of 
 magnetic monopoles with charge $\alpha_j$ and antimonopoles with charge $- \alpha_{j+1}$.  
 The  net effect of bions can be  encoded  into a {\it prepotential}, out of which we may 
 derive the  potential.   Interestingly enough, the relation between the prepotential and potential 
 is the  same as the relation between the superpotential and potential in $\N=1$ SYM, 
 modulo the absence of  the Higgs scalar in the former (where it is massive).  
 We give  the analytic derivations of characteristic
  sizes of chromoelectric flux tubes  in QCD(adj) in 
  the small $S^1$ regime. 
 
Let us complete the introduction by saying that  closer and deeper inspection of non-supersymmetric 
QCD-like theories may also be used to build the relation between the inner goings-on of 
the  supersymmetric and nonsupersymmetric gauge  theories. Suffice it to say that, the integrable systems 
which emerges in the QCD(adj) are  variants of the affine Toda systems 
\cite{Corrigan:1994nd, Hollowood:1992by, Kneipp:2007fg}, which also appeared in the 
discussions of  $\N=1,2$ SYM, and elliptic curves  \cite{D'Hoker:2002pn}.     
This direction will not be explored in this paper, but is potentially interesting. 
\footnote{
There are also recent, interesting works  on the dynamics of four-dimensional gauge theories, 
in particular  for pure Yang-Mills,  see  \cite{Diakonov:2007nv, Tomboulis:2007iw, Lenz:2007st}, and for 
lattice works, see \cite{DiGiacomo:1999fa, Greensite:2003bk} and references therein. Also,  good 
reviews covering different aspects of monopoles and instantons can be found in \cite{Tong:2005un,  
Weinberg:2006rq, Schafer:1996wv}.}

\section{Dynamics of  $SU(2)$  QCD(adj) on small $S^1 \times \R^3$ }
\label{sec:dyn}
\subsection{Perturbation theory}
\label{sec:per}
First, we wish to give the microscopic derivation  of the dual theory  \ref{Eq:dQCD}. 
The action of  $SU(N)$ QCD(adj) defined on $\R^3 \times S^1$   is 
\begin{equation}
S= \int_{\R^3 \times S^1} \frac{1}{g^2} \tr \left[ \frac{1}{4} F_{MN}^2 + i \bar \lambda^I \bar 
\sigma^M D_{M} \lambda_I \right]
\label{eq:Lagrangian}
\end{equation}
where $\lambda_I= \lambda_{I,a} t_a,   a=1,  \ldots, N^2-1 $ is  the Weyl fermion in adjoint representation, 
$F_{MN}$ is the nonabelian gauge field strength, and  $I$ is the flavor index, and the generators are normalized as $\tr \;  t^a t^b = \delta^{ab}$. 
The  
classical theory  possesses a  $U(n_f)$ flavor symmetry  whose $U(1)_A$ part is anomalous. 
The symmetry of the  quantum theory  is 
\begin{equation}
(SU(n_f) \times \Z_{2N n_f})/\Z_{n_f}
\end{equation}
The quantum 
theory has the  dynamical strong scale $\Lambda$, which arises via dimensional transmutation, 
and  is given  by $\Lambda^{b_0} = \mu^{b_0} e^{-8 \pi^2/g^2(\mu)N }$ where $\mu$ is the 
renormalization 
group scale and $b_0= (11-2n_f)/3$.  We consider small $n_f  $ so that asymptotic freedom is preserved.  The $n_f=1$ case (SYM) will be discussed  separately.   We first discuss 
$N=2$   QCD(adj), and $N\geq 3$ will  be discussed in section \ref{sec:sun}.

At small $S^1$ $(L \Lambda \ll1)$,  due to asymptotic freedom,  
the gauge coupling is  small and a  perturbative Coleman-Weinberg analysis  is 
reliable \cite{Coleman:1973jx}.  Let $U({ x}) = P e^{i\int dx_4 A_4({x}, x_4)}$  be the path-ordered 
holonomy of the spatial Wilson line wrapping the $S^1$, and sitting at the point 
${x} \in \R^3$.  
Integrating out the heavy KK-modes  along the $S^1$ circle, $|\omega_n|\geq \omega_1$ where
$\omega_n= \frac{2 \pi}{L}n, n \in \Z$, induces a nontrivial effective potential for $U({x})$ 
\cite{Unsal:2006pj}.  
\begin{equation}
V^{+}[U] = (-1+ n_f) \frac{2}{\pi^2 L^4} \sum_{n=1}^{\infty} \frac{1}{n^4}  |\tr U^n|^2 
\label{pot1}, \qquad  U(x) = P e^{i\int dx_4 A_4({x}, x_4)} \equiv e^{i L \Phi}
 \end{equation}
 Note that the stability of the center symmetry is induced by massless adjoint fermions with periodic boundary conditions. In this sense, this theory {\it does not} require the double-trace deformations  to achieve phases of unbroken center symmetry 
 \cite{Unsal:2008ch,Myers:2007vc,Ogilvie:2007tj}.  
The action for the classical zero-modes reduce to
\begin{eqnarray}
S= \int_{\R^3} \frac{L}{g^2} \tr  &\Big[ \frac{1}{4} F_{\mu \nu}^2 + \frac{1}{2} (D_{\mu} \Phi)^2 +  
g^2 V(| \Phi |) 
+i \bar \lambda^I (\bar \sigma^{\mu} D_{\mu} + \bar \sigma_4 [ \Phi, \; ]) \lambda_I \Big]
\label{Eq:compact}
\end{eqnarray}
The minimum of the potential $V_{\rm eff}$ is located at $L|\Phi| = \frac{\pi}{2}$, 
hence 
\begin{equation}
 U= \left( \begin{array}{cc} 
  e^{i \pi/2} & \\ 
  & e^{-i \pi/2}
  \end{array}
  \right) \qquad  {\rm or}  \;\;\;\;  \Phi= \left( \begin{array}{cc} 
  \pi/2  & \\ 
  & - \pi/2
  \end{array}
  \right)
 \end{equation}
Since $\tr U = 0$, the $\Z_2$
center symmetry is preserved. By the   Higgs mechanism, the
gauge symmetry is broken down as  
\begin{equation}
SU(2) \rightarrow  U(1) 
\end{equation}
Because of  adjoint Higgs mechanism, the neutral fields aligned with $U$ along  the Cartan subalgebra ($A_{3,\mu}, \lambda^{I}_{3}$)   remain massless, and off-diagonal components acquire mass, given by the separation between the eigenvalues of the Wilson line 
\begin{equation}
m_{W^{\pm}}= m_{\lambda^{I, \pm}} =  \pi/L
\end{equation}
where $\pm$ refers to the charges under unbroken $U(1)$. Therefore, in perturbation theory,  
the low energy theory is a $d=3$  dimensional abelian $U(1)$ gauge theory 
with $n_f$  massless fermions with a free action 
 \begin{equation}
S= \int_{\R^3 } \frac{L}{g^2} \left[ \frac{1}{4} F_{3,\mu \nu}^{2} + i \bar \lambda^I_{3} \bar 
\sigma^\mu \partial_{\mu} \lambda_{3,I} 
\right]
\label{eq:pert}
\end{equation}

\begin{figure}[t]
{
  \parbox[c]{\textwidth}
  {
  \begin{center}
  \includegraphics[width=3in]{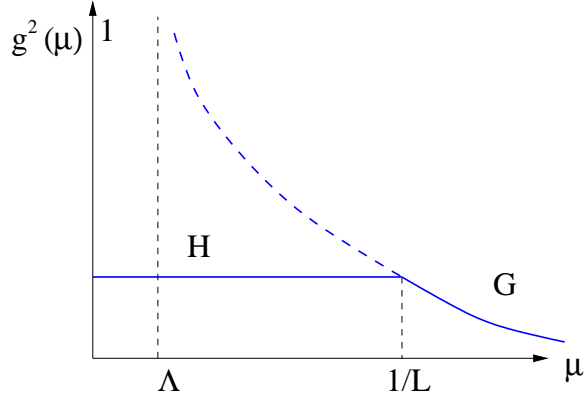}
  \caption
    {%
Summary of perturbative analysis: Solid line indicates the running of the gauge coupling in QCD(adj) compactified on a small circle $S^1$ with circumference $L$, and dashed line is the usual running on $\R^4$.   
In the regime  $1/L \gg \Lambda $   perturbative Coleman-Weinberg analysis is reliable,  and 
leads to a radiatively induced gauge symmetry breaking   $G \rightarrow H$ where $G=SU(2)$ and 
$H= U(1)$.   To all orders in perturbation theory, the long-distance theory described  by $H$ is free due to absence of charged massless  excitations.  This is reminiscent of 
the  $\N=2$  SYM theory on $\R^4$, for which gauge symmetry breaking takes place on the semi-classical domain of the  moduli space. 
    }
  \end{center}
  }
\label{fig:flow}
}
\end{figure}

At distances shorter than $L$, the coupling constant flows 
according to the four-dimensional renormalization group.
Since the heavy   $W^{ \pm}, {\lambda^{I, \pm}}$ which are charged under 
$U(1)$  decouple from the long-distance physics at scale $L$ and above,   the coupling 
  constant ceases to run  at $1/L \gg \Lambda$  much before the  strong coupling sets in,  see 
  Fig.\ref{fig:flow}. In perturbation  theory, this is the whole story.

\subsection{Nonperturbative effects and abelian duality}
\label{sec:non}
Nonperturbatively, the  perturbatively free infrared fixed point  is unstable.   This follows from the 
existence of monopoles (strictly speaking, these are monopole-instantons or fractional instantons), at the cores of which the $U(1)$  symmetry  of the free theory enhances to 
the whole non-abelian $SU(2)$. 

Because of  gauge symmetry breaking via a {\it compact} adjoint Higgs field, there are two types of monopoles, BPS and KK,  
as well as their antimonopoles ${\overline {\rm BPS}}$,   ${\overline {\rm KK}}$ 
\cite{Lee:1997vp, Kraan:1998pm, Kraan:1998sn,Lee:1998bb}.
\footnote{Were the gauge symmetry broken by a noncompact Higgs field, the KK monopole would not be there. As we will discuss, 
this is the case in the extension of the Polyakov model in the presence of adjoint fermions, a theory which does not confine.} These four types of monopoles are distinguished by their quantized 
magnetic and topological     charges  $\Big(   \int F , \int F \widetilde F  \Big)$ normalized  as 
\begin{eqnarray}
&{\rm BPS}: (+1, \half), \qquad   & {\overline {\rm BPS}}:(-1, -\half), \ \cr 
&{\rm KK}\; : (-1,  \half),  \qquad  &{\overline {\rm KK}}\; :(+1, -\half).    
\end{eqnarray}
Because of  the chiral anomaly  relation \cite{'tHooft:1976fv}, 
\begin{equation}
\partial_{M} J^{M 5} = \frac {g^2 (2Nn_f)} {32 \pi^2}  \tr  F_{MN} {\widetilde F}^{MN} 
\end{equation}
each  object with a non-vanishing topological charge is associated with a certain number of fermionic zero-modes.  Integrating both sides over the space, we find  
\begin{equation}
\Delta Q_5 =n_\lambda - n_{\bar \lambda}= 4n_f \int  \frac{g^2}{32\pi^2}   
\tr  F_{MN} {\widetilde F}^{MN} =   \left \{
\begin{array}{ll}
  4n_f  (\frac{1}{2})=2 n_f  &\qquad  {\rm  \; for \; BPS \; or \;  KK} \\
    4n_f(-\frac{1}{2})=-2 n_f  &\qquad  {\rm  \; for \; {\overline {\rm BPS}} \; or \;   {\overline {\rm KK }}} \\
\end{array}
\right .
\label{Eq:index}
\end{equation}
where the term inside the parenthesis is the topological charge.  As it should be clear, $4n_f$ is the 
number  of fermionic zero-modes associated with a four-dimensional  instanton,  
whose topological charge is $+1$. 
Since the topological charges of monopoles are a fraction of the one of the instanton, they are sometimes referred as fractional instantons. Clearly, a BPS-KK pair has the correct quantum numbers to be the constituents of the instanton \cite{Kraan:1998sn,Lee:1998bb}. 

 By abelian duality \cite{Polyakov:1976fu, Deligne:1999qp}, we know that the functional integral in a  gauge theory in the presence of a  single monopole with charge $\pm 1$ 
located at the position $x$ is 
equivalent to the insertion of an operator $e^{\pm i \sigma(x)}$ in the path integral of the dual theory. 
However, the index version of the chiral anomaly relations 
\ref{Eq:index} tells us that a monopole acts as it contains a source for every fermion flavor, and an antimonopole acts as if it contains a sink for  all fermion flavors. 
Adapting  a combination of techniques developed  by 't Hooft  \cite{'tHooft:1986nc}  
and  by  Polyakov  \cite{Polyakov:1976fu} 
 to our problem, we can sum up all the monopole effects. 
The   functional integral (with a source) in the presence of a monopole 
\begin{equation} 
\int DA_{\mu} D\psi^{I}  D \bar \psi^{I} e^{-S_{\rm one \; mon.}(A, \psi, \bar \psi) + J \det \psi^{I} \psi^{J} +
 \bar J \det \bar \psi^{I} \bar \psi^{J} } 
\end{equation}
is  the same as having 
\begin{equation} 
e^{-S_0}\int D\sigma  D\psi^{I}  D \bar \psi^{I} e^{-S_{d, 0} (\sigma, \psi, \bar \psi)+ J \det \psi^{I} \psi^{J} +
 \bar J \det \bar \psi^{I} \bar \psi^{J} }  e^{i \sigma(x)} \det_{I, J} \psi^{I} \psi^{J}
\end{equation}
where $S_{d, 0} (\sigma, \psi, \bar \psi)= \int_{\R^3} \left[ 
\half (\partial \sigma)^2  + i \bar \psi^{I}  
\gamma_{\mu} \partial_{\mu}  \psi_{I}  \right] $ is the free kinetic term. 
Hence,  a functional integral in the 
presence of a monopole can be translated into having a monopole  vertex $e^{i \sigma(x) }$ with 
accompanying  fermionic zero-modes. 
We can insert the monopole at any $x \in \R^3$,  and we can consider an arbitrary number of them.  
 The sum  over all possible monopole configurations is  
\begin{eqnarray}
&&\sum_{n_{\rm BPS}=0}^{\infty}  \sum_{n_{\overline{\rm BPS}}=0}^{\infty}  
\sum_{n_{\rm KK}=0}^{\infty}  \sum_{n_{\overline{\rm KK}}=0}^{\infty} 
\frac{ e^{- (n_{\rm BPS} + n_{\overline {\rm BPS}}+ 
 n_{\rm KK}+  n_{\overline {\rm KK}})S_0   }}{ n_{\rm BPS} ! \; n_{\overline{\rm BPS}}! \;  n_{\rm KK}! \; 
  n_{\overline{\rm KK} }! }
   \left[ \int d^3 x e^{i \sigma(x)} \det_{I, J} \psi^{I} \psi^{J} \right]^{n_{\rm BPS}} \cr &&\cr && \cr
    &&  \left[\int d^3 x e^{-i \sigma(x)} \det_{I, J} \bar \psi^{I} \bar \psi^{J} \right]^{n_{\overline{\rm BPS}}} 
         \left[ \int d^3 x e^{-i \sigma(x)} \det_{I, J} \psi^{I} \psi^{J} \right]^{n_{\rm KK}} 
            \left[ \int d^3 x e^{i \sigma(x)} \det_{I, J} \bar \psi^{I} \bar \psi^{J} \right]^{n_{   \overline{\rm KK}}} \qquad 
\end{eqnarray}
Performing the summation yields monopole induced terms of order  $e^{-S_0}$ in our  effective lagrangian 
\begin{equation} 
\exp\Big[ \int d^3x  \; e^{-S_0}  (e^{i \sigma} + e^{-i \sigma})    
( \det_{I, J} \psi^I \psi^J +     \det_{I, J} \bar \psi^I \bar \psi^J ) \Big]
\end{equation}
Therefore, the  combined effect of BPS and KK monopoles is   $\cos \sigma  \det \psi^I \psi^J$.
This vertex is manifestly invariant under continuous $SU(n_f)$ flavor symmetry, acting as 
$\psi \rightarrow U \psi$ where $U \in SU(n_f)$.  The microscopic theory 
also possesses a   $\Z_{4n_f}$ discrete 
chiral symmetry. \footnote{ \label{fn:chiral}More generally,  consider $SU(N)$ QCD(adj) with $n_f$ flavors.  The chiral symmetry is  $[SU(n_f) \times \Z_{2N n_f}] / \Z_{n_f} $,  where the  common  $\Z_{n_f}$ is  factored out to prevent double counting.  
The  $\Z_2$ subgroup of the $\Z_{2N}$ is $(-1)^F$ fermion number modulo $2$, which cannot be spontaneously broken so long as Lorentz symmetry is unbroken.  
Thus, the only genuine  discrete chiral symmetry of $SU(N)$ QCD(adj)  which may potentially be 
broken is the  remaining $\Z_N$, irrespective of the number of flavors.  In small $S^1$, we explicitly demonstrate the existence of $N$ vacua, and spontaneous breaking of chiral $\Z_N$ symmetry (which is intertwined with the discrete shift symmetry of photon).    
This $\Z_N$ symmetry should not be confused with the spatial center symmetry, $\G_s= \Z_N$
which is  unbroken in spatial compactification of QCD(adj). 
}
The effective theory, in order to respect the $\Z_{4n_f}$ discrete chiral symmetry, intertwines it with a  
discrete shift symmetry of the dual photon:  
\begin{equation}
\psi^I \rightarrow e^{i  2\pi/(4n_f)} \psi^I, \;\;\;\;     \sigma \rightarrow \sigma + \pi 
\end{equation}
both of which  acts as  negation  on   the   determinantal fermion vertex and cosine combinations
\begin{equation}
 \det_{I, J} \psi^I \psi^J \rightarrow -  \det_{I, J} \psi^I \psi^J ,  \qquad 
 \cos \sigma \rightarrow - \cos \sigma
 \end{equation} 
respectively, so that the effective theory respects the real symmetries of the underlying theory.  
 
\begin{figure}[t]
{
  \parbox[c]{\textwidth}
  {
  \begin{center}
  \includegraphics[width=5in]{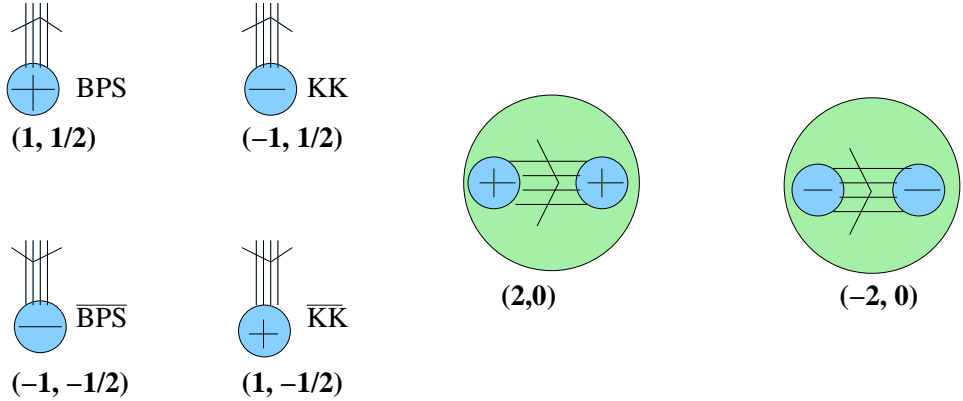}
  \caption
    {%
(Left)Magnetically and topologically charged monopoles carry compulsory fermion zero-modes. Consequently, they cannot induce a bosonic potential for the dual photon.    (Right) Topologically null, magnetically charged bions have no  external fermionic legs. Hence, they induce the leading  bosonic potential, which implies  mass  for the  dual photon and confinement. The figure is for $SU(2)$ with $n_f=2$. 
The combination of the BPS- KK monopoles (which is not depicted)  is an instanton (or caloron). It is present in confined phase, but is not the source of the dual photon mass term. 
    }\label{fig:monopole}
  \end{center}
  }
}
\end{figure}

In the effective Lagrangian,  this is the set of all nonperturbative effects at order  $e^{-S_0}$ in the  $e^{-S_0}$ expansion.  
However, the  discrete $\Z_2$ shift symmetry   $\sigma \rightarrow \sigma + \pi$,  
unlike a continuous shift symmetry, cannot prohibit a mass term for the scalar $\sigma$. Clearly,  
a term $e^{-S_0} \cos \sigma$ is forbidden  by $\Z_2$.  But its square is an allowed operator. 
If fermions were not present, 
\begin{equation}
 e^{-S_0} \cos \sigma \sim  e^{-S_0} (e^{i \sigma} + e^{-i \sigma} )
 \end{equation} 
  would be an allowed term as in the Polyakov's 
discussion of the  Georgi-Glashow model, and would induce a mass term of order $e^{-S_0/2}$ for dual photon.   However, 
because of the index theorem \ref{Eq:index},  a monopole must come with fermion zero-modes, and a term such as 
$e^{i \sigma}$ cannot appear on its own, but must appear in combination 
  $e^{i \sigma} \det_{I, J} \psi^I \psi^J  $.

Symmetry principles also tell us that, at the $e^{-2S_0}$ order,    we can write 
\begin{equation}
[e^{-S_0} 
\cos \sigma ]^2 \sim e^{-2S_0} ( 1+ 1+ e^{2i \sigma}  +e^{-2i \sigma} )
\end{equation}
and this would  generate a mass term for the dual photon, hence leading  to  confinement.   
We wish to understand the dynamical origin of this potential.  

Let us first forget about the issues about fermion zero-modes,  and decide on the basis 
of quantum numbers, which objects may contribute to the nonperturbative  potential.  
Since we know that, due to index theorem, such an object can not be a monopole, let us enlist all possible pairs of monopoles,  
the   magnetic and topological charges of constituents and pairs, and  the  types of the 
long range Coulomb interactions, repulsive or attractive. 
In nonsupersymmetric QCD(adj)  with $2\leq n_f \leq 4$, the list of all Coulomb interaction channels 
for monopoles  is given by 
\begin{equation} 
 \begin{array}{|l|l|c|l|}
 \hline
      {\rm Type}  & {\rm Type} &   \sigma {\rm -int.} &  
      \left( \int F, \int F \widetilde F \right)  \\
      \hline
   {\rm BPS}- e^{+i \sigma}  &  {\rm BPS}-e^{+i \sigma}  & {\rm rep.} &   (+1, +\half) +  (+1, +\half) =    
   (2,    1)  \\ 
{\rm BPS} &  \overline {\rm BPS}- e^{-i \sigma}  & {\rm att.} &  (+1, +\half) +  (-1, - \half) =   ( 0, 0)   \\ 
{\rm BPS} &  {\rm KK }-e^{-i \sigma}  & {\rm att. } &  (+1, +\half) +  (-1, +\half) =   ( 0 , 1)  \\
{\rm BPS} &  \overline {\rm KK}-e^{+i \sigma}   & {\rm rep.} &  (+1, +\half) +  (+1, -\half) =   (2, 0)  \\
{\overline  {\rm BPS}}  &  {\overline {\rm BPS}} & {\rm rep.} &   (-1, -\half) +  (-1, -\half) =    
   (-2,   -1)  \\   
{\overline {\rm BPS}} &   {\rm KK} & {\rm rep.} &  (-1, -\half) +  (-1, + \half) =   (-2, 0)   \\
{\overline{\rm BPS}} & { \overline {\rm KK }}  & {\rm att. } &  (-1, -\half) +  (+1, -\half) =   (0 , -1)  \\
{\rm KK} & {\rm KK}  & {\rm rep.} &  (-1, +\half) +  (-1, +\half) =   (-2,  1)  \\
  {\rm KK}  &  {\overline {\rm KK}}   & {\rm att.} &   (-1, +\half) +  (+1, -\half) =    
   (0,    0)  \\
{\overline {\rm KK} }  &  \overline {\rm KK} & {\rm rep.} &  (+1, -\half) +  (+1, - \half) =   (2, -1) 
\\ 
\hline 
          \end{array} 
\end{equation}

In the presence of the fermion zero-modes,  
 the (bosonic) potential must arise due to the  sector of the theory with  zero topological charge so  that  there will not be any  fermion zero-mode insertions in it.  In other words, the objects which may contribute to the potential must be topologically indistinguishable from the perturbative vacuum.  
 
 This immediately rules out the four possible monopoles,  and six of the ten pairs  in our list from 
 contributing to the bosonic potential.   In particular,  the two identical  monopole  configuration 
 such as BPSBPS    with  $(1, \half) + (1, \half)= (2, 1)$
 have  the correct magnetic charge, but its topological charge does not permit it to contribute to the  
 bosonic potential.  Another interesting combination 
 which does not lead to the  confining potential  is 
 a BPSKK pair. 
The BPSKK pair in fact constitute an instanton (sometimes called a caloron, 
\cite{Kraan:1998sn,Lee:1998bb} ) with charge  $(1, \half) + (-1, \half)= (0, 1)$ 
 and does not induce mass term for the dual photon. 
  
The monopole and antimonopole pairs such as 
BPS-${\overline {\rm BPS}}$  are topologically null, but also magnetically neutral.  Their contribution to 
the effective potential can only be an uninteresting  constant. 
There remains a single option: a bound state of  BPS monopole, and KK antimonopole,  
  BPS-$\overline {\rm KK}$, and its conjugate.   
   Such an  object has the correct   quantum numbers $(\pm1, \half) + (\pm1, -\half)= (\pm2, 0)$.
   We referred to  this object as a  magnetic bion, see  Fig.\ref{fig:monopole}. 
 Consequently, 
 the bion is  the prime candidate  which may
lead  to confinement in QCD(adj) in the $L\Lambda \ll 1$ regime.   

However, there is an immediate puzzle with this proposal.  There is a long range  Coulomb repulsion 
between  BPS-${\overline {\rm KK}}$ constituents of the bion.  If we wish to have a bound state, there must exist   an attractive interaction which overcomes the repulsive Coulomb force.  Happily, there is! 

\subsection{Pairings and attractive multi-fermion exchanges }
\label{sec:pai}
The presence of fermion zero-modes  changes things drastically.
We will demonstrate that   for the pairs with net topological charge zero, 
there exists  an attractive $V_{\rm eff} \sim \log r$  interaction  between the constituents due to fermion pair  exchanges.    
For the pairs with a  nonvanishing topological charge,   the constituents do not interact at all due to chirality at leading order.   
  
Let us first show the first assertion:  Consider  BPS  and ${\overline {\rm KK}}$ monopoles located 
at    $ x,  y  \in \R^3$, where $|x-y| \gg 1$. ($x, y$ are dimensionless coordinates in units of $L$.) 
 We can extract their interactions  from the 
 connected correlator of the BPS    vertex $V_{\rm BPS}(x) $,  and $ { \overline {\rm KK}  }$ vertex  
 $ V_{ \overline {\rm KK}  } (y) $  
  in the free dual theory with action  $S_{d,0}(\sigma, \psi, \bar \psi)$
\begin{eqnarray}
&& \langle V_{\rm BPS} (x)   V_{ \overline {\rm KK}  } (y) \rangle_0 =  \langle 
 e^{i  \sigma(x) }  \det_{IJ}{\psi^I  \psi^J}(x)   
 e^{+ i \sigma(y) }  \det_{I'J'}{\bar \psi^{I'}  \bar \psi^{J'}} (y) \rangle_0  \cr
 && =  \langle 
 e^{i  \sigma(x) }   e^{ i  \sigma(y) }  \rangle_0 \; 
\langle  
 \det_{IJ}{\psi^I  \psi^J}(x)   
  \det_{I'J'}{ \bar \psi^{I'}   \bar \psi^{J'}} (y) \rangle_0 \cr 
&&  \sim e^{- G(x-y)}   [S_F(x-y)]^{2n_f}
  \end{eqnarray}
  where $G(x-y)=\frac{1}{4 \pi |x-y|}$ is the Coulomb potential, which is the 
  position  space propagator of the $\sigma$  field,  $G(x) = \int \frac{d^3p}{(2 \pi)^3} e^{ipx} \frac{1}{p^2}$  and 
  $S(x)= \frac{\sigma^{\mu} x^{\mu}}{4 \pi |x|^3}$  
  is the $d=3$ dimensional free fermion propagator 
  $S(x)=   \sigma^{\mu}\frac{\partial}{\partial x_{\mu}} G(x)$. 
  The static interaction potential between the BPS and  $ { \overline {\rm KK}  }$ pair is 
\begin{eqnarray}
V_{\rm eff}(x-y)= - \log  \langle V_{\rm BPS} (x)   V_{ \overline {\rm KK}  } (y) \rangle_0  = 
  \frac{1}{4 \pi |x-y|} + 4n_f \log |x-y| 
\end{eqnarray}
Asymptotically, $4 n_f \log |x-y| $ is the dominant attractive interaction term, and it easily overcomes 
the Coulomb repulsion. Therefore, there exist a stable bion bound state with the total magnetic and topological  charge $(+2, 0)$, and antibion with charge $(-2, 0)$. 
It should be noted that 
the stability of the magnetic bion relies on the masslessness (or lightness)  of the adjoint fermions. In this case,  the fermion induced attraction overcomes the Coulomb repulsion for a small  range of (light) fermion mass. For more details, see section \ref{YM*}. 

 It should also be noted that similar fermion zero-mode induced pairings of topological excitations were discussed earlier in the literature  by Callan, Dashen, and Gross and others  \cite{Callan:1977gz, Schafer:1996wv}  in the context of instantons  on $\R^4$. The  pairing mechanism is similar to what we have found above, in that case instanton and anti-instanton form  molecules  due to  attraction induced by fermions. 
 Interestingly, the form of the attractive interaction is the same both in $\R^4$  and $\R^3 \times S^1$, and is a logarithmically attractive interaction proportional to the number of flavors, $n_f \log r$.  As  these instanton--anti-instanton molecules are  magnetically and topologically neutral, they play no role in confinement and chiral symmetry realization in the small $S^1$ regime of QCD(adj). In our topological semi-classical expansion, they appear at order  $e^{-2NS_0}$ and are  a negligible effect.   

Analogously,  the net interaction between a  BPS-${\overline {\rm BPS}}$ pair  is attractive in both 
interaction channels, either   Coulomb, or  fermion exchange interactions. The long-distance  attraction has the form 
$ -\log \langle V_{\rm BPS} (x)   V_{ \overline {\rm BPS}  } (y) \rangle_0 = -\frac{1}{4 \pi |x-y|} + 4n_f \log |x-y|$.  

Because of  chirality of the  underlying theory,  the interaction between 
pairs with the same  topological charge vanishes identically:
$
 \langle V_{\rm BPS} (x)   V_{{ \rm BPS}  } (y) \rangle_0 =  \langle V_{\rm BPS} (x)   V_{ {\rm KK}  } (y) \rangle_0 =   0$.

Since the topological charge of the   magnetic bion is zero, it does not have any fermion zero-mode attached to it.  Since magnetic bions and  antibions have   $\pm 2$ magnetic charges, they will 
lead to Debye phenomena.   
The appropriate effective potential induced by bions is indeed what we wrote based on  symmetry arguments:
\begin{equation}
V(\sigma)= [e^{-S_0}  \cos \sigma ]^2 \sim e^{-2S_0} ( 1+ 1+ e^{2i \sigma}  +e^{-2i \sigma} )
\end{equation}
The terms in the potential have  an interpretation as the contribution  of respectively 
 ${\rm BPS}{\overline {\rm BPS}}   +  {\rm KK}{\overline {\rm KK}} +  {\rm BPS}{\overline {\rm KK}}+ {\rm KK}{\overline {\rm BPS}}$. 
 
More precisely, the interaction terms in the lagrangian are due to monopole and bion contributions.  
The monopole contributions necessarily involve the  fermion interactions.  Schematically, the nonperturbatively induced interaction terms will always be 
\begin{equation} 
L_{\rm int}  = \underbrace{\sum_{\rm bions}  V_{\rm bion}}_{\int F \widetilde F =0 }  + 
\underbrace{\sum_{\rm monopoles}  V_{\rm monopole}}_{\int F \widetilde F =\pm \half } 
\end{equation}
Therefore, the dual QCD lagrangian  for $SU(2)$ QCD(adj) on small $S^1 \times \R^3$ 
is given by  
\begin{eqnarray}
L^{\rm dQCD} = \frac{1}{2} (\partial \sigma)^2 -  b\;  e^{-2S_0}\cos 2 \sigma  
+  i \bar \psi^I \gamma_{\mu} \partial_{\mu} \psi_I   
 + c \; e^{-S_0}  \cos \sigma   
( \det_{I, J} \psi^I \psi^J +  \rm c.c.) \qquad 
\end{eqnarray}
up to higher-order (insignificant) terms in $e^{-S_0}$.

The potential term for the dual photon, when expanded around one of its two minima (located at $0$ 
and $\pi$),  provides a mass term for the dual photon.  From  the point of view of Euclidean field theory, the photon mass 
 is the inverse  Debye screening length in the plasma of magnetic bions. 
  On a fixed time slice of a timelike Wilson loop,  the inverse photon mass   
is the  
thickness  of the chromoelectric  flux tube formed between two external electric test charges.  
Just like the Polyakov model \cite{Polyakov:1976fu} on $\R^3$,  the QCD(adj) on small 
$S^1 \times \R^3 $ exhibits linear confinement, 
\begin{equation}
V_{\rm linear} (R) \sim  e^{-S_0} R,  
\end{equation}
and the potential energy of a pair of the electric source separated by a distance $R$ grows linearly with separation.  

{\bf Remark:}  The results and approach  of this work should not be confused 
with 't Hooft's abelian projection scheme \cite{'tHooft:1981ht}, which only leaves an $U(1)^{N-1}$ gauge symmetry.    Hence, monopoles in that case are gauge artifacts, which is fine in the prescribed gauge. 
  In our case, 
the gauge symmetry breaking   $SU(N) \rightarrow U(1)^{N-1}$  is dynamical, and 
is a well-controlled effect due to the radiatively induced Coleman-Weinberg 
potential.   The QCD(adj) in the $ L \Lambda \ll 1 $ regime tells us that, in the presence of fermions, 
the idea of monopole condensation no longer holds due to fermion zero-modes. Despite this fact, 
the qualitative and beautiful  idea 
of dual superconductivity  of 't Hooft and Mandelstam  \cite{'tHooft:1981ht, Mandelstam:1974pi}    
is still realized at a quantitative level,  albeit  via condensation of the 
pairs with combined magnetic and topological charges $(\pm 2, 0)$.

As emphasized, the presence of monopoles is not sufficient to induce confinement, or monopole condensation.     Better appreciation of the above picture can come with the study of a 
 Yang-Mills Higgs system with adjoint fermions on $\R^3$,  a system with monopoles and yet 
 no confinement.

\subsection{Noncompact Higgs with adjoint fermions on  $\R^3$,  and the lack  of confinement}
\label{sec:nonc}
Affleck, Harvey and Witten   studied extensions of  Polyakov's model in the presence of an 
adjoint Dirac fermion  on $\R^3$   \cite{Affleck:1982as}.  The generalization of their argument to multiple 
flavors is obvious.    They analyzed (among other things) a Yang-Mills Higgs system 
which  possesses the   same  action as Eq.\ref{Eq:compact}, except the fact that 
the compact adjoint Higgs field in   Eq.\ref{Eq:compact} 
is   substituted by a non-compact one.   
\begin{equation} 
V_{\rm eff}^{\rm compact}(|\Phi|) \rightarrow  V_{\rm eff}^{\rm noncompact}(|\Phi|) 
\label{Eq:PolyakovE}
\end{equation}
Since the chiral anomaly is absent in odd dimensions, the noncompact 
model has a genuine $U(n_f)$ symmetry  whose $U(1)$ part is fermion number.  
Ref.\cite{Affleck:1982as} showed quite explicitly that such a  model does  {\bf not} confine.
 Photons remain  at infinite range nonperturbatively, and it is indeed the Goldstone boson of 
  the spontaneously broken $U(1)$ fermion number symmetry. 
    Their arguments are  essentially based on symmetries, and  index theorem   by Callias   
    \cite{Callias:1977kg}, and  explicit zero-mode construction by  Rebbi and Jackiw 
  \cite{Jackiw:1975fn}. Here, we wish to provide 
    a  simple dynamical explanation for this phenomena. 
  
  Since gauge symmetry breaking occurs via a {\it noncompact} adjoint Higgs field, there is no longer a 
  KK monopole. Thus, in order to obtain the long-distance effective action from our discussion  in previous section, 
    we must delete all KK monopole related terms from our effective action. 
  Hence, the interaction lagrangian is $L_{\rm int} \sim V_{\rm BPS} + V_{\overline {\rm BPS}} +  
  V_{{\rm BPS}\overline {\rm BPS}} $. Consequently, 
  \begin{equation} 
L_{\rm eff}^{\rm noncompact} = \frac{1}{2} (\partial \sigma)^2 + i \bar \psi^I \gamma^{\mu} \partial_{\mu} \psi_I + 
a e^{-S_0}( e^{i  \sigma} \det \psi^I \psi^J + {\rm c.c.})
\end{equation} 
where we ignored a trivial cosmological constant which may be induced by a BPS$\overline {\rm BPS}$ pair.  This is indeed the generalization of  Ref.\cite{Affleck:1982as} to multiflavor $(n_f >1)$.   
The effective action  is respectful to all the symmetries of the underlying
theory, in particular $SU(n_f) \times U(1)$ symmetry, where the former is manifest.
  The $U(1)$ fermion number symmetry acts as 
\begin{equation}
\psi^I \rightarrow e^{i  \alpha} \psi^I, \;\; \bar \psi^I \rightarrow e^{-i  \alpha} \bar \psi^I, \;\; 
 \sigma \rightarrow \sigma -  2n_f \alpha \; .
\end{equation}
and  prohibits any kind of mass term (or potential) for the dual photon.  This is the symmetry 
which breaks down spontaneously, and the dual photon   is the Goldstone boson. 
  
Clearly, the    only topologically neutral object (which may contribute to the bosonic potential)  is BPS$\overline {\rm BPS}$ pair. But such an 
  object has vanishing magnetic charge. Since there are no  topologically null, but   magnetically charged 
  carriers  in the vacuum of the model studied in   \cite{Affleck:1982as},  the Debye mechanism is not possible. Hence, the  photon remains infinite range nonperturbatively. 
  The inability to form magnetically charged bions is 
  the dynamical reason for the absence of confinement in the extension of Polyakov's model in the presence of adjoint fermions. 
 
 This discussion also shows  that the presence of monopoles in the Yang-Mills Higgs systems with adjoint fermions is a necessary  but insufficient condition to have confinement.  
  In particular, it also exhibits that, in such systems, 
 condensations of objects with non-vanishing topological charge (monopole condensation) 
 do not occur.

\subsection{Magnetic bions in $\N=1$ SYM on small $S^1 \times \R^3$ }
\label{sec:mag}
The  generalization of the  discussion in section \ref{sec:pai}  to $SU(2)$ 
 $\N=1$ supersymmetric gauge theory is easy, yet important.  
All one needs to take care  of is an extra massless scalar  which remains massless in perturbation theory.  Hence it  should be incorporated into long-distance physics. 
 With the inclusion of the $\phi$-scalar,  the monopoles may interact via  $\phi$-exchange, $\sigma$-exchange and fermion pair exchange channels:
 \begin{equation} 
 \begin{array}{|l|c|c|c|c|c|} 
 \hline
      {\rm Type}  & {\rm Type} &   \sigma {\rm -int} & {\phi {\rm  -int.}} & {\rm combined} & 
             \left( \int F, \int F \widetilde F \right)  \\ \hline
   {\rm BPS}- e^{-\phi+ i \sigma}  &  {\rm BPS}-  e^{-\phi+ i \sigma} 
   & {\rm rep.} &  {\rm att.}  & 0&   (1, \half) + (+1, +\half) = (2, 1)  \\
{\rm BPS} &  \overline {\rm BPS}-  e^{-\phi- i \sigma}   & {\rm att.} &   {\rm att.}  & 2( {\rm att.} ) &
 (1, \half) + (-1, -\half) = (0, 0) 
  \\
{\rm BPS} &  {\rm KK }-  e^{+\phi- i \sigma}  & {\rm att.} & {\rm rep.} & 0 &  (1, \half) + (-1, +\half) = (0, 1) \\
{\rm BPS} &  \overline {\rm KK} -  e^{+\phi+ i \sigma}  & {\rm rep.} &  {\rm rep.} & 2({\rm rep.}) 
&  
 (1, \half) + (+1, -\half) = (2, 0) 
 \\        
 \ldots & \ldots & \ldots & \ldots &\ldots &\ldots \\ \hline
   \end{array} 
\end{equation}
Incorporating the scalar field $\phi$ into monopole operators, we find 
\begin{eqnarray}
&& {\rm BPS}: e^{ -\phi + i \sigma} \psi \psi,  \qquad {\rm KK}: e^{ +\phi - i \sigma}   \psi  \psi,  \cr 
&& {\rm \overline {BPS}}: e^{- \phi -i \sigma}  \bar \psi \bar \psi,  \qquad {\rm \overline{KK}}:
  e^{ + \phi + i \sigma}   \bar \psi \bar \psi,  
\end{eqnarray}
The bosonic potential is due to the  sector of the theory with net zero topological charge, so 
that  there will not be any  fermion zero-mode insertion in it.  
 Thus
\begin{eqnarray}
{\rm BPS}\overline{\rm BPS}  +   {\rm KK}\overline{\rm KK} +  
{\rm BPS}\overline{\rm KK} +  {\rm KK}\overline{\rm BPS} &&=
e^{-2S_0} ( e^{ -2\phi} +  e^{ +2\phi}-  e^{  i 2  \sigma}  -  e^{ -2 i \sigma} ) \cr 
&&= e^{-2S_0} | e^{z} - e^{-z}|^2
\end{eqnarray}
where we defined $z= -\phi+ i \sigma$.  Remarkably, 
 the magnetic  bions  already know  that there is an underlying  superpotential,  
given by 
 \footnote{
Strictly speaking,  this superpotential is the form acquired after the superHiggs mechanism. }    
\begin{equation}
{\cal W}(z)= e^{-S_0} (e^z+ e^{-z})
\end{equation}
The 
long-distance effective action for SYM on small $S^1 \times \R^3$  is 
\begin{eqnarray}
L_{\rm eff}^{\rm SYM} = &&\half (\partial \sigma)^2 + \half (\partial \phi)^2 -  \; c^2 e^{-2S_0}
(\cos 2 \sigma -
 \cosh 2 \phi)  \cr  \cr 
&&+  i \bar \psi \gamma_{\mu} \partial_{\mu} \psi   
 + c \; e^{-S_0}\left[   (e^{ -\phi + i \sigma} +  e^{+ \phi - i \sigma} ) \psi \psi  +   
  (e^{ -\phi - i \sigma} +  e^{+\phi + i \sigma} ) \bar \psi \bar \psi   
 \right]  
  \qquad 
  \label{Eq:SYMc} 
\end{eqnarray}
The $\Z_{2N}= \Z_{4}$ discrete chiral symmetry of the original theory is also manifest in the 
effective theory
\begin{equation}
\psi^I \rightarrow e^{i  2\pi/4} \psi^I, \;\;\;\;     \sigma \rightarrow \sigma + \pi 
\end{equation}
This symmetry breaks down spontaneously  to $\Z_2= (-1)^F$   where $F$ is fermion number, leading to  
the appearance of two isolated vacua. 

The dynamics of the $\N=1$ SYM on $\R^3 \times S^1$  is  previously analyzed  
by imbedding it 
into F theory in  Ref. \cite{Katz:1996th}, 
 and by using the elliptic curves of $\N=2$ SYM
combined with  the mass deformation in \cite{Seiberg:1996nz}.  
The  works of Davies et.al 
\cite{Davies:1999uw, Davies:2000nw}
provided a  clear field theory exposition of the nonperturbatively induced 
effects in such theories.
The general strategy of these papers was  to calculate the monopole  operator  first, then use supersymmetry as a completion device to find the superpotential, 
 hence bosonic potential. 
For fermionic terms, our strategy  is the same as in these earlier works.  
For the bosonic potential, our strategy 
is different.  Rather then using  supersymmetry as a completion tool to derive bosonic potential, we preferred to  delineate on its 
microscopic (physical)  origin.  In essence, we   identified  topologically null configurations which are topologically indistinguishable from the perturbative vacuum, and hence can contribute to the potential. 
Summing  up their contributions gives us the bosonic  potential, which can also  be derived from the superpotential.  

These two approaches in the case of $\N=1$ SYM are identical. The latter approach has a higher value in our opinion due to the fact that it does not make any  
reference to supersymmetry, and  works for non-supersymmetric  QCD-like theories. 
Our analysis  makes it manifest that  the mechanism of confinement in $\N=1$ SYM  is not   monopole condensation,  i.e.,  condensation 
of excitations with topological charge $\pm \half$,  rather of objects with 
topological charge $0$. This physical fact was not understood in earlier important works 
on the subject \cite{Katz:1996th, Seiberg:1996nz, Davies:1999uw, Davies:2000nw}. 
Up to our knowledge, our work is the first analytic demonstration of confinement induced by 
non-self-dual topological excitations. Needless to say, even  the  issue of presence or absence of such  topological excitations  was not discussed. 
  We conclude this section by pointing out that the mechanism of the 
confinement    in supersymmetric  $\N=1$ SYM is same as the one in 
 nonsupersymmetric QCD(adj) theories  in the $L\Lambda \ll 1$ regime, both of which are    
 magnetic bion condensation, a new class of (non-self-dual) topological excitations.

 In the dimensional reduction of $\N=1$  SYM  down to $\R^3$, confinement does not occur as 
 shown in \cite{Affleck:1982as}.   The distinctions are so important that   it is worthwhile rederiving their 
 results following the consideration of this paper, and   
 explaining  the absence of confinement on dynamical grounds. 

\subsection{The $\N=2$ SYM on $\R^3$  and lack  of confinement, again}  
\label{sec:SYM}
    Delete 
all  the terms in the effective action \ref{Eq:SYMc}  which are  related to KK monopole. 
(This is the same statement as the $\phi$ field becomes noncompact on the  $\R^3$ limit.)
  This leaves us with 
BPS and $\overline {\rm BPS}$ induced operators (involving fermion bi-linears)  and a BPS$\overline {\rm BPS}$ induced term in the bosonic potential in the lagrangian \ref{Eq:SYMc}: 
\begin{eqnarray}
L_{\rm eff}^{\rm n.c.} = \half (\partial \sigma)^2 + \half (\partial \phi)^2 -  \; c^2 e^{-2S_0}
 e^{-2 \phi}   
+  i \bar \psi \gamma_{\mu} \partial_{\mu} \psi   
 + c \; e^{-S_0}\left[   e^{- \phi + i \sigma}  \psi \psi  +   
  e^{ -\phi - i \sigma} \bar \psi \bar \psi   
 \right]  
  \qquad 
  \label{Eq:SYMnc} 
\end{eqnarray}
which is the same as the lagrangian in  \cite{Affleck:1982as}.
The $\Z_{2N}$ discrete chiral symmetry of SYM on locally four-dimensional settings elevates to the full 
$U(1)$ fermion number on $\R^3$ due to absence of chiral anomaly in odd dimensions. 
  The continuous $U(1)$ symmetry acts as 
\begin{equation}
\psi^I \rightarrow e^{i  \alpha} \psi^I, \;\; \bar \psi^I \rightarrow e^{-i  \alpha} \bar \psi^I, \;\; 
 \sigma \rightarrow \sigma -  2  \alpha \; .
\end{equation}
and  prohibits any kind of explicit mass term (or potential) for the dual photon.  This is the symmetry 
which breaks down spontaneously, and the dual photon   is the Goldstone boson. 
The runaway potential $e^{- 2 \phi}$  does not have a  vacuum at  finite $\phi$. 

On dynamical grounds,  the absence of confinement is due to  the inability to form long range magnetic  bions in SYM vacuum on $\R^3$. The BPS$\overline {\rm BPS}$ pairs are neutral, 
 and the photon remains  infinite range in a medium of neutral molecules. In other words, it remains massless nonperturbatively as demanded from a Goldstone particle,
  and this implies the absence of confinement. 
 
\section{ $SU(N)$ QCD(adj), bions,  and secret integrability? }
\label{sec:sun}
The $SU(N)$ QCD(adj) theory undergoes  gauge symmetry breaking on sufficiently small spatial 
$S^1$ due to a perturbative Coleman-Weinberg potential.  The gauge symmetry breaking is 
$SU(N) \rightarrow U(1)^{N-1}$.   For simplicity,  we will add a decoupled "center of mass"   
degree of freedom to the original theory and consider gauge symmetry breaking of the form 
$U(N) \rightarrow U(1)^N$.  This is a technical trick, and in the spontaneously broken gauge theory,
the center of mass mode  decouples from the dynamics. Hence, our goal is to determine 
the dynamics of the $N-1$  modes $\frac{U(1)^N}{U(1)_{\rm c.m. }}$
 
The monopoles may be described by their magnetic charges, topological charge and their  action. 
 The magnetic charges  of the  $N$ types of (BPS and  KK)  monopoles  
 under  unbroken gauge symmetry $U(1)^N$    are  proportional to the simple roots and affine root 
  of the  Lie algebra, respectively. 
 The simple roots are given by 
 \begin{eqnarray}
&\alpha_1&= (1, -1, 0, \ldots, 0)= e_1-e_2 \cr
&\alpha_2&= (0, 1, -1, , \ldots, 0)=e_2 -e_3  \cr
&\alpha_i&= (0, \ldots, 1, -1,  \ldots 0)=e_i -e_{i+1} \cr
&\ldots  & \cr 
&\alpha_{N-1}&= (0, \ldots,, 0,  1, -1)=e_{N-1} -e_{N} 
\label{Eq:basis}
\end{eqnarray}
 and the affine root is   
\begin{equation}
\alpha_N \equiv  - \sum_{j=1}^{N-1} \alpha_j =  (-1, 0, 0, \ldots ,1)= e_N- e_1
\end{equation}
It is convenient to define  the simple  $\Delta^{0}$ and 
affine (extended)  $\Delta^{0}_{\rm aff}$ root systems of the   associated  
Lie algebra: 
\begin{equation}
 \Delta^{0} \equiv \{ \alpha_1, \alpha_2, \ldots , \alpha_{N-1} \}, \qquad  
\Delta^{0}_{\rm aff} \equiv \{ \alpha_1, \alpha_2, \ldots , \alpha_{N-1},  \alpha_N \},  
\end{equation}
The latter is the one relevant for QCD(adj) on $\R^3 \times S^1$.  More generally, in the Yang-Mills Higgs systems with adjoint fermions,  if the Higgs field is noncompact, the monopole and antimonopole  charges are valued in  $\Delta^{0}$ and,   $- \Delta^{0}$, respectively. 
 If the Higgs field is compact, then there is an extra monopole, and the charges take 
values in  $\pm \Delta^{0}_{\rm aff}$.  

The topological charges $\int F \tilde F$ are correlated with the sign of the two sets  $\pm \Delta^{0}_{\rm aff}$.  Thus, the quantized magnetic and topological charges are 
\begin{equation} 
\int_{S^2} F^i  =   \pm \frac{2 \pi}{g} \alpha^i, \qquad  
\int F \widetilde F \equiv
\frac {g^2} {32 \pi^2}
\int   \tr  F_{MN} {\widetilde F}^{MN} =
\pm \frac{1}{N}, 
\end{equation}
The action of a monopole with charge $\alpha_i$ and   topological charge  
$
\int F \widetilde F=  \pm \frac{1}{N}
$
is given by 
$S_{0, i}= \frac{8 \pi^2}{g^2}\int F \widetilde F = \frac{8 \pi^2}{g^2 N}$. 
 Because of  the presence of the  effective potential for the Wilson line,  the monopoles 
of QCD(adj) theory (except for $n_f=1$ which is supersymmetric)    
do  not saturate  the BPS bound. But the correction are perturbative in $g^2$ 
and we will neglect them.  

The long-range Coulomb interaction of monopoles (in the absence of fermions) is given by 
\footnote{We  set $\frac{2\pi}{g}$ to unity as in our discussion of $SU(2)$ to lessen the clutter 
in expressions.  All physical quantities are measured in units of $L$, which is also set to unity. 
We will restore both  quantities if necessary.}   
 \begin{equation}
V ( \alpha_i,  \pm   \alpha_j, r) = \frac{\alpha_i.  (\pm \alpha_{j}) }{4 \pi r} = 
\pm \frac{2 \delta_{ij} - \delta_{i, j +1}   -  \delta_{i, j -1}}{4 \pi r}, \qquad i, j=1, \ldots N
 \label{Eq:Dynkin}
\end{equation}
which translates to self and nearest neighbor  interaction between monopoles  in the Dynkin space. 
The inner product of the roots of the associated Lie algebra  is a basis independent statement, though the above choice of the basis    \ref{Eq:basis}  is due  to its   visual simplicity.  

We are  now ready to generalize the  
derivation of effective potential for $SU(2)$  QCD(adj) to  $SU(N)$ with $1< n_{f} \leq 4$. 
Our discussion  will be brief.   

Were the adjoint fermions absent, a  monopole with charge $\alpha_{j}$  would be  associated 
with  operator   $e^{i \alpha_j \sigma}$. 
Because of index theorem \ref{Eq:index},  any object with a nonvanishing topological charge $ (1/N)$
 must have   $\Delta Q_{5} = 2n_{f}$ fermions attached to it.   As discussed in footnote \ref{fn:chiral}, 
 the underlying QCD(adj) theory has  $[SU(n_f) \times \Z_{2Nn_f}]/ \Z_{n_f}$ continuous and discrete chiral symmetries.  
 The manifestly  $SU(n_f) $ invariant  fermion vertex with $2n_f$ fermion insertion is given by 
  $ \det_{IJ}{\alpha_i \psi^I  \alpha_i \psi^J} $
 where the  determinant is over the  flavor index.  
 Here, we use a vector  notation
\begin{equation}
\sigma= (\sigma_{1}, \ldots, \sigma_{N}), \qquad  \psi^{I}= (\psi^{I}_{1}, \ldots,  \psi^{I}_{N}), \qquad 
\alpha_{i}\sigma= \sigma_{i}- \sigma_{i+1} 
\end{equation}
As stated earlier, the center of mass mode is extraneous and decouples from the dynamics completely.  
Hence,  the appropriate monopole and antimonopole  operators  are 
\begin{eqnarray}
 V_{\alpha_i}  =    e^{i \alpha_i \sigma }  \det_{IJ}{\alpha_i \psi^I  \alpha_i \psi^J} , \qquad 
   V_{- \alpha_i}  =    e^{-i \alpha_i \sigma }  \det_{IJ}{\alpha_i \bar \psi^I  \alpha_i \bar \psi^J}  
\end{eqnarray}
This means, the interaction Lagrangian at $O(e^{-S_{0}})$ is given by 
\begin{equation}
e^{ -S_{0}  } \sum_{\alpha_{i} \in \Delta_{\rm aff}^{0}}  \left(  e^{i \alpha_i \sigma }  \det_{IJ}{\alpha_i \psi^I  \alpha_i \psi^J}  +   e^{-i \alpha_i \sigma }  \det_{IJ}{\alpha_i \bar \psi^I  \alpha_i \bar \psi^J}  \right)
\end{equation}
This vertex is invariant under   $(SU(n_{f}) \times \Z_{2Nn_{f} }   )/\Z_{n_{f}}$ as desired.  
 The discrete chiral  symmetry acts as 
\begin{equation}
\psi^I \rightarrow e^{i  2\pi/(2Nn_f)} \psi^I, \;\;\;\;     \bar \psi^I \rightarrow e^{-i  2\pi/(2Nn_f)} \bar 
\psi^I, \qquad 
\sigma \rightarrow \sigma - \frac{2 \pi}{N} \sum_{j=1}^{N-1} \mu_{k} 
\label{inter}
\end{equation}
where 
$\mu_{k}$ are the $N-1$ fundamental weights (not the weight of fundamental representation) 
of the associated Lie algebra.     They are defined by the reciprocity relation, 
\begin{equation}
\frac{2 \alpha_{i} \mu_{j}}{\alpha_{i}^{2}}= \alpha_{i} \mu_{j}= \delta_{ij}
\end{equation}
The shift in the photon field is called the Weyl vector, and we will often abbreviate it as 
\begin{equation}
\rho \equiv \sum_{j=1}^{N-1} \mu_{j}, \qquad  {\rm such  \;  that}  \qquad e^{i \frac{2 \pi}{N} \rho \alpha_j} =  e^{i \frac{2 \pi}{N} }, \qquad j=1, \ldots, N
\end{equation}
The action of the discrete chiral symmetry on $SU(n_f)$ singlets is a   $\Z_N$  symmetry transformation,  
\begin{equation}
\det_{I, J} \alpha_{i}\psi^I \alpha_{i} \psi^J \rightarrow e^{{i2 \pi}/N} \det_{I, J} \alpha_{i}\psi^I  \alpha_{i}\psi^J, \qquad e^{i \alpha_i \sigma } 
\rightarrow e^{-{i2 \pi}/N} 
 e^{i \alpha_i \sigma } \; .
\end{equation}
Consequently,  the monopole induced interaction terms (which are of  order $e^{-S_{0}}$)  are 
respectful the discrete (and continuous)  symmetries of the underlying theory. 

Exactly as in the $SU(2)$ discussion, this is the net effect of the topologically nontrivial 
 sector of the  theory which saturates the lagrangian at  order  $e^{-S_{0}}$.   
In particular,  a would-be (confining) potential term for the $\sigma$ field 
 \begin{equation}
e^{ -S_{0}  } \sum_{\alpha_{i} \in \Delta_{\rm aff}^{0}}  \left( 
 e^{i \alpha_i \sigma }  +   e^{-i \alpha_i \sigma}  \right)
\end{equation}
 is forbidden by the $\Z_N$ shift symmetry  $\sigma - \frac{2 \pi}{N} \sum_{j=1}^{N-1} \mu_{k} $
 of the dual photon.    This is a consequence of having adjoint fermions in the system. In the absence 
 of fermions, such as  a pure Yang-Mills compact Higgs system,    this {\it is} the leading term which renders 
 all the photons massive, with masses of order $e^{-S_{0}/2}$. We will see that  in QCD(adj), the 
 masses of photons are  of order $e^{-S_{0}}$, and there is a  $\Z_{N}$ shift symmetry respecting potential  at order $e^{-2S_{0}}$.

\subsection{Attractive channels,  bions, and a prepotential} 
 \label{sec:att}
We must examine the  combinations of the monopole-antimonopole pairs 
with magnetic  charges from the two sets   $\Delta_{\rm aff}^{0}$ and  $-\Delta_{\rm aff}^{0}$ with respective topological charges $\frac{1}{N}$ and   $-\frac{1}{N}$.  Because of the presence of many possible pair
that one can construct, this  may {\it a priori} seem arbitrary.  However, the theory does something 
remarkable. At order $e^{-2S_{0}}$, the fermion zero-mode exchanges  only pairs the monopoles with charge $\alpha_{j}$ with their  nearest neighbor antimonopoles, with charges $- \alpha_{j\pm 1}$ in the Dynkin space. These combinations  are  the magnetic bion states.    
(There are also neutral  monopoles and antimonopole pairing of the same kind, but the magnetic charge of such an object is zero and not so 
interesting in nonsupersymmetric QCD(adj). It has an effect in SYM as discussed in section 
\ref{sec:mag}.)

Let us first find  the attractive channels.  We can extract the interaction of a monopole with charge 
 $\alpha_{i} $ and antimonopole with charge $-\alpha_{j}$ by inspecting   its connected correlator in the 
functional integral of the free theory with the action  $S_{d, 0}(\sigma, \psi, \bar \psi)$. 
\begin{eqnarray}
&& \langle V_{\alpha_i} (x)   V_{- \alpha_j} (y) \rangle_0 =  \langle 
 e^{i \alpha_i \sigma(x) }  \det_{IJ}{\alpha_i \psi^I  \alpha_i \psi^J}(x)   
 e^{-i \alpha_j \sigma(y) }  \det_{I'J'}{\alpha_j \bar \psi^{I'}  \alpha_j \bar \psi^{J'}} (y) \rangle_0  \cr
 && =  \langle 
 e^{i \alpha_i \sigma(x) }   e^{-i \alpha_j \sigma(y) }  \rangle_0 \; 
\langle  
 \det_{IJ}{\alpha_i \psi^I  \alpha_i \psi^J}(x)   
  \det_{I'J'}{\alpha_j \bar \psi^{I'}  \alpha_j \bar \psi^{J'}} (y) \rangle_0 \cr 
&&  \sim e^{+ \alpha_i. \alpha_j G(x-y)} (\alpha_i \alpha_j)^{2n_f} [S_F(x-y)]^{2n_f}
  \end{eqnarray}          
The connected correlator is only nonzero if $\alpha_{i}\alpha_{j}$ is nonzero, and induces 
 a  logarithmic binding potential of the form 
\begin{equation}
V_{\rm eff}(x-y)=  
\left \{
\begin{array}{ll}
   + \frac{1}{4 \pi |x-y|} + 4n_f  \log |x-y| 
    &\qquad  {\rm  \; for} \; i= j \pm 1 \\
     - \frac{2}{4 \pi |x-y|} + 4n_f  \log |x-y|  &\qquad  {\rm  \; for} \;  i=j   
      \\
\qquad   \qquad     0 & \qquad {\rm \; otherwise} \; .
\end{array}
\right .
\label{Eq:effective2}
\end{equation}
 If $i=j$, then both Coulomb and fermion zero-mode exchange induced forces are attractive. 
If $i=j\pm1$, then the Coulomb interaction is repulsive, but the attractive fermion exchange term easily dominates.

Now, we are ready to define the magnetic bions in the spontaneously broken 
$SU(N)$ gauge theory. 
A bion is a  bound state of the monopole associated with magnetic charge $\alpha_i$ and antimonopole associated  with charge $- \alpha_{i+1}$ with null topological charge:
 \begin{eqnarray}
 Q_i= \alpha_i - \alpha_{i-1}= 2e_i - e_{i+1} - e_{i-1}, \qquad  
 \int   F \widetilde F=  0
 \qquad i=1, \ldots N
\end{eqnarray}
Restoring the prefactors and writing more explicitly, the magnetic bion (antibion) charges are given 
under the $U(1)^N$   gauge group as 
 \begin{eqnarray}
 Q_i=\pm   \frac{2 \pi} {g} \Big( 0, \ldots, \underbrace{-1}_{i-1}, \underbrace{2}_{i},\underbrace{-1}_{i+1}, 
 \ldots , 0 \Big) 
\end{eqnarray}
This means, bions interact via a next-to-nearest neighbor interaction in the Dynkin space:  
For high-rank gauge groups ($N\geq5$), 
\begin{eqnarray}
 Q_i Q_j= 6 \delta_{ij}  - 4\delta_{i,j+1} -   4\delta_{i,j-1} +   \delta_{i,j+2}  + \delta_{i,j-2}, \qquad N \geq 5 
\end{eqnarray}
In order to find the bion-bion interactions in low-rank gauge groups $N\leq 4$, we need to identify  
nodes $j\equiv j+N$  in the  (affine) Dynkin diagram as there are less than five nodes. Consequently,
\begin{eqnarray}
&& Q_i Q_j= 6 \delta_{ij}  - 4\delta_{i,j+1} -   4\delta_{i,j-1} +   2\delta_{i,j+2} , \qquad N=4 \cr 
 && Q_i Q_j= 6 \delta_{ij}  - 3\delta_{i,j+1} -   3\delta_{i,j-1}  , \qquad \qquad \qquad  N=3 \cr 
  && Q_i Q_j= 8 \delta_{ij}  - 8\delta_{i,j+1} , \qquad \qquad \qquad \qquad \qquad N=2 
   \end{eqnarray}
The long-range interactions of magnetic bions are given by Coulomb's potential and are equal to  
  \begin{equation}
V ( Q_i,  \pm   Q_j, r) = \frac{Q_i.  (\pm Q_{j}) }{4 \pi r} = 
\pm \frac{  6 \delta_{ij}  - 4\delta_{i,j+1} -   4\delta_{i,j-1} +   \delta_{i,j+2}  + \delta_{i,j-2}}
{ 4 \pi r}
\label{Eq:Dynkin2}
\end{equation}
The meaning of this formula is clear. Two magnetic bions   with  charges $(Q_i, Q_i)$   repel,    $(Q_i, Q_{i\pm 1})$   attract,  $(Q_i, Q_{i\pm 2})$   repel, and no interactions for pairs $(Q_i, Q_{i+k})$ with $k>2$.   
The overall sign of the interactions is reversed for the bion-antibion pairs. 

Now, we can convert the Coulomb gas of magnetic bions into a field theory following Polyakov's treatment \cite{Polyakov:1976fu}. We only quote the result, since the manipulations are standard. 
 The operator   appropriate for a bion molecule located at $x \in \R^{3}$ is 
\begin{equation}
e^{i Q_{i} \sigma(x)}= e^{i\alpha_{i} \sigma(x) }  e^{- i\alpha_{i-1} \sigma(x) }  
\end{equation}
Clearly, this is manifestly invariant under the $\Z_{N }$ shift symmetry of the photon which acts as 
$e^{i\alpha_{i} \sigma(x) } \rightarrow e^{-i 2 \pi/N}  e^{i\alpha_{i} \sigma(x) } $.  The bosonic effective potential is a sum over all bion and antibion contributions given by 
 \begin{equation}
V(\sigma)= - e^{ -2S_{0}  } \sum_{i=1}^{N}   \left( 
 e^{i Q_i \sigma }  +   e^{-i Q_{i} \sigma}  \right) = -2e^{ -2S_{0}  } \sum_{i=1}^{N}   
 \cos{Q_i \sigma }  
\end{equation}
There is something  remarkable about this potential, in fact surprising.  
It can be derived from a  {\it prepotential}, just like a bosonic potential in the supersymmetric system may be derived from a superpotential.  In order to see this, rewrite the potential 
$V(\sigma) $ as 
 \begin{eqnarray}
V(\sigma) &&=    - e^{-2S_0} \sum_{i=1}^{N}   \left( 
 e^{i \alpha_i \sigma }    
 e^{- i \alpha_{i-1} \sigma }  + e^{- i \alpha_i \sigma }    e^{ i \alpha_{i-1} \sigma } 
 \right)   \cr
 &&= e^{-2S_0} \sum_{i=1}^{N}  \;   | 
 e^{i \alpha_i \sigma } -  e^{ i \alpha_{i-1} \sigma } |^{2} +{ \rm constant} 
 \end{eqnarray}
 where constant is unimportant.  Define the prepotential as
 \begin{equation}
{\cal W}(\sigma)= e^{-S_0} \sum_{ \alpha_{i} \in \Delta_{\rm aff}^{0}  } e^{  i \alpha_i \sigma } \; . 
\end{equation}
Hence, the potential may be written as  
\begin{eqnarray}
V(\sigma) =   \sum_{i=1}^{N}    \Big| \frac{\partial {\cal W} }{\partial \sigma_i} \Big|^2 = 
 e^{-2S_0} \sum_{i=1}^{N} 
| e^{i \alpha_i \sigma } -  e^{i \alpha_{i-1} \sigma }|^2, \qquad {\rm QCD(adj)} \; n_f>1
\end{eqnarray}

The reader familiar with the supersymmetric affine Toda theories  will recognize the form of our 
(nonsupersymmetric) prepotential as the superpotential.  
In order to describe the infrared of $\N=1$ SYM on small $S^1$,  one must incorporate the extra massless scalars into the potential: All one needs to do is a holomorphic completion of our formula. 
Not surprisingly,  
\begin{eqnarray}
V(z, \bar z ) =   \sum_{i=1}^{N}    \Big| \frac{\partial {\cal W} }{\partial z_i} \Big|^2 =  
 \sum_{i=1}^{N} 
| e^{i \alpha_i z } -  
e^{ i \alpha_{i-1} z }|^2 \; , \qquad {\rm SYM} 
\end{eqnarray}
The fact that the potential can be derived from a prepotential as above implies that the  classical equations of motions for the $\sigma$ field can be reduced to a first-order one.  

Let us finalize this section by writing the final form of the dual of the QCD(adj) lagrangian on small 
$S^1 \times \R^3$ with $1< n_f \leq 4$ flavors: 
\begin{eqnarray}
&& L^{\rm dQCD} = \half (\partial \sigma)^2 -  
b\;  e^{-2S_0} \sum_{\alpha_{i} \in \Delta_{\rm aff}^{0}}  
| e^{i \alpha_i \sigma } -  e^{+ i \alpha_{i-1} \sigma }|^2
 \cr 
&&+  i \bar \psi^I \gamma_{\mu} \partial_{\mu} \psi_I   
 + c \;  e^{ -S_{0}  } \sum_{\alpha_{i} \in \Delta_{\rm aff}^{0}}  \left(  e^{i \alpha_i \sigma }  \det_{IJ}{\alpha_i \psi^I  \alpha_i \psi^J}  +   e^{-i \alpha_i \sigma }  \det_{IJ}{\alpha_i \bar \psi^I  \alpha_i \bar \psi^J}  \right)
\label{Eq:dQCD2} 
\end{eqnarray} 
The dualQCD  lagrangian and the physics it encapsulates, which will be discussed next, 
 are  the essential result of this paper. 

\subsection{Brief comparison to deformed YM theory}
\label{YM*}
In this section, we will briefly outline the main difference between the deformed YM theory (to be  abbreviated  as YM*) studied in \cite{Unsal:2008ch} and QCD(adj).   In YM*,  
due to the absence of fermionic matter, the monopole operators do not carry any fermionic zero-modes. Thus, the dual theory can  be obtained by summing over all monopole operators.   The dual description of YM* theory is 
 \begin{equation}
L^{\rm dYM^*}=  \half (\partial \sigma)^2  - e^{ -S_{0}  } \sum_{\alpha_{i} \in \Delta_{\rm aff}^{0}}  \left( 
 e^{i \alpha_i \sigma }  +   e^{-i \alpha_i \sigma}  \right)
\end{equation}
 On the other hand,  for QCD(adj) with massless fermions, the dual description is given in \ref{Eq:dQCD2}.  In particular, in QCD(adj),  monopole operators do not contribute a mass gap for the dual photons. 
 The bosonic potential which renders the dual photon massive  is effectively the square of the potential given for YM*. If we just write the dual of the gauge sector for QCD(adj), the difference is more transparent. 
  \begin{equation}
L^{\rm dQCD}=  \half (\partial \sigma)^2  - e^{ -2S_{0}  } \sum_{\alpha_{i} \in \Delta_{\rm aff}^{0}}  \left( 
 e^{i (\alpha_i - \alpha_{i-1}) \sigma }  +   e^{-i (\alpha_i - \alpha_{i-1}) \sigma}  \right)
\end{equation}
Consequently, the functional form of the mass gap for gauge fluctuations 
 is different in these two class of theories as it will be compared in \ref{sec:mas}.

If one keeps center symmetry stable and turn on a mass term for the adjoint fermion, 
the magnetic bion induced confinement mechanism should be replaced by magnetic monopole induced confinement. In particular, 
for heavy  adjoint fermions, the index theorem on $S^1 \times \R^3$ does not apply. Thus, 
the theory must reduce to YM*. 

The more interesting case is the  light 
adjoint fermions.  In principle, the fermion zero-modes may be (softly) lifted by the mass term. 
As a result, the modified monopoles operators will also contribute to mass gap and confinement. For sufficiently light fermions, the bion mechanism dominates. It would be interesting to examine the transition from magnetic bion induced confinement to magnetic monopole induced confinement in more detail in future work.

\subsection{The vacuum structure of QCD(adj)}
\label{sec:vac}
The bosonic potential of nonsupersymmetric QCD(adj) has $N$ gauge inequivalent isolated vacua, 
aligned along the Weyl vector $\rho$  
\begin{equation}
\sigma=  \{ 0, \frac{2\pi}{N}, \frac{4 \pi}{N}, \ldots ,  \frac{(N-1) 2 \pi}{N} \}   \rho 
\end{equation}
in the field space. This is the same as $\N=1$ SYM studied in \cite{Davies:2000nw}.  
 Since each component of $\sigma$ is a periodic variable with periodicity $2\pi$, 
there exists a 
physical  congruence between $\sigma$ and $\sigma'$ which is  separated by an element of the root lattice $\Lambda_{\rm r}$. 
\begin{equation}
\sigma \equiv \sigma + 2 \pi \alpha  \qquad {\rm for \; some } \;   \alpha \in \Lambda_{\rm r}
\end{equation} 
Since the sum of all fundamental weights is a root,  
$ \rho =  \sum_{j=1}^{N-1} \mu_j  \in  \Lambda_{\rm r}$, this implies there only exist $N$ gauge inequivalent vacua when the (global) gauge symmetry redundancies  
are removed. Let us the abbreviate and label the vacuum states in Hilbert space as  
   \begin{eqnarray}
&&  |\Omega_{\frac{2 \pi  k}{N} \rho  +  \Lambda_{\rm r}} \rangle \equiv     |\Omega_k\rangle  \equiv 
    |\Omega_{k+N} \rangle    , \qquad k=0, \ldots N-1  
   \qquad   \cr
&& {\rm Ground \; states}= \Big\{  |\Omega_0 \rangle,    \; \;    |\Omega_1 \rangle,  \ldots,  |\Omega_{N-1} \rangle \Big\} 
    \end{eqnarray}    
which form a one-dimensional representation of $\Z_N$ shift symmetry, (which is  intertwined with $\Z_N$ discrete chiral symmetry, see footnote \ref{fn:chiral}.) This means, the (large) physical Hilbert space spits into  $N$ superselection sectors, each of which may be built upon the associated vacuum.  
The choice of the vacuum breaks the $\Z_N$ discrete chiral symmetry (which is same as $\Z_N$ shift symmetry of the dual photon)  spontaneously.  Note that QCD(adj) also possess a $\G_s= \Z_N$ spatial center symmetry which remains unbroken regardless of the size of the $S^1$, and
which should not be confused with  the $\Z_N$ axial or equivalently,   $\Z_N$ shift symmetry of dual photons.

\subsubsection{Mass gap in the gauge sector}  
\label{sec:mas}
The small fluctuations 
around one of the $N$  minima of the $- \cos Q_i\sigma$  potential 
shows  that the $N-1$ dual photon acquires masses   proportional to $e^{-S_0}$.  
In order to see this, let us expand the  nonperturbative bion induced potential  to quadratic order in dual photon $\sigma$
\begin{eqnarray}
V(\sigma_i) &=& - e^{-2S_0} \sum_{i=1}^{N}   \cos Q_i\sigma = - e^{-2S_0} \sum_{i=1}^{N} 
\cos (2{ \sigma}_i- {\sigma}_{i+1} -  {\sigma}_{i-1} ) \cr  
&= &
\half e^{-2S_0} \sum_{i}  \left( 6{ \sigma}_i^2- 4 \sigma_i{\sigma}_{i+1}  -4 \sigma_i \sigma_{i-1} +   {\sigma}_{i}  \sigma_{i+2}  + {\sigma}_{i}  \sigma_{i-2}  \right)  \qquad {\rm bion \;  induced} \;\;\;
 \end{eqnarray}
 If  the fermions were absent, and the gauge symmetry was still broken by a compact adjoint Higgs field as in YM* \cite{Unsal:2008ch},  the quadratic fluctuations would be described by the  nearest neighbor coupled  harmonic oscillator 
 \begin{eqnarray}
V(\sigma_i) &=& 
\half e^{-S_0} \sum_{i}  \left( 2{ \sigma}_i^2-  \sigma_i{\sigma}_{i+1}  - \sigma_i \sigma_{i-1}\right) 
\qquad {\rm  monopole \;  induced, YM^*}
 \end{eqnarray}
which is not the case in QCD(adj).  The bion induced  ``hopping'' terms are  
next-to-nearest neighbor and of order $e^{-2S_0}$ as opposed to the monopole induced hopping terms which are just nearest neighbor, and of order $e^{-S_0}$.

The quadratic fluctuations can be diagonalized   
 by  using the discrete Fourier transform   $\sigma_p =  \frac{1}{\sqrt N}  \sum_{j=0}^{N-1}
  \omega^{jp} \sigma_j $  in Dynkin space:
\begin{eqnarray}
V( \sigma_p ) &=&\half e^{-2S_0} \sum_p  ( 6 -4 \omega^{-p} -4  \omega^p + \omega^{-2p}  +  \omega^{2p}  )
\;\;  \sigma_p  \sigma_{-p}  \cr 
&=&    \half e^{-2S_0}
 \sum_p  ( \omega^{p/2} -   \omega^{-p/2} )^4   
\;\;  \sigma_p  \sigma_{-p} 
=  \half           e^{-2S_0}                \sum_{p} ( 2 \sin \frac {p \pi}{N})^4   \;\;  \sigma_p  \sigma_{-p} \qquad 
\end{eqnarray}
Restoring the dimensions, we obtain the mass spectrum of the $N-1$ dual photons  as   
\begin{equation}
m_p^{\rm QDC(adj)} \sim \Big( \Lambda(\Lambda L)^{b_0 -1} =  \Lambda(\Lambda L)^{(8-2n_f)/3} \Big) \times 
 (2 \sin \frac {p \pi}{N})^2,  \qquad p=1, \ldots, N-1  
\label{Eq:Spec}
\end{equation}
This result implies that   the gauge sector of the QCD(adj) theory is quantum mechanically gapped due to non-perturbative effects, and    permanently confines 
external electric charges at small $S^1\times \R^3$ limit.  Note that the analogous formula for the mass gap in the gauge sector of YM* is given by 
\begin{equation}
m_p^{\rm YM^*} \sim  \Lambda(\Lambda L)^{5/6} 
 \sin \frac {p \pi}{N},  \qquad p=1, \ldots, N-1  
\label{Eq:Spec2}
\end{equation}
in the  $\Lambda L  \ll 1$ regime.

The masses are graded according to the  $\Z_N$ center group of $SU(N)$ in one to one 
correspondence with the representations  ${\cal R}_p$ of 
$SU(N)$ under the center group.   There are two equivalent physical interpretation for the mass gap: 
one as the inverse  Debye screening length  in a magnetic conductor (in a Euclidean setting), and the other 
is  the inverse thickness of the chromoelectric flux tubes in a magnetic superconductor 
(at a fixed time in a  Minkowski setting).  (See \cite{Callan:1977gz} for a parallel  discussion in the context of the  Polyakov model.)

 Imagine a large, planar Wilson loop in a representation with charge $p$ under the center  $ \Z_N$, \   In the small $S^1$ regime (where gauge symmetry is broken to the abelian subgroup),  we may regard the Wilson loop as carrying an electric current along the contour of the loop. Hence, by Maxwell's equation, the current  generates a magnetic field along the axis perpendicular to the plane of the loop, within the boundary $C$ of the loop surface $\Sigma$.  
   The external magnetic field   cannot penetrate  into the magnetic conductor above a penetration depth, due to Debye  screening. 
 The mobile magnetic  charge carriers (bions)     form a dipole layer in the vicinity of the surface $\Sigma$ to   prevent the penetration of the external magnetic field into the magnetic conductor, which is the vacuum of QCD(adj) from   Euclidean viewpoint.  The thickness of the dipole layer for the Wilson loop 
 with $\Z_N$ charge $p$ is the inverse of the photon mass $m_p^{-1}$.
 
We may visualize a Wilson loop at a fixed time slice.   This is a system with 
  $\pm p$ $\Z_N$  chromoelectric  sources located at two boundaries  of the 
 fixed time slice  of the  Wilson loop.   
 There exist   a stable chromoelectric flux tube in between the two. 
   Since   the dual superconductor  expels the electric field,  the   flux  lines are trapped within tubes with quantized    flux.  
 The $N-1$ classes of the photon  masses are indeed the    
inverse characteristic sizes of the $N-1$ types of the  chromoelectric 
flux tubes, both of which  are  a class function of the $\Z_N$ center group.  
 In a weakly coupled  regime, making $L$ larger    reduces the thicknesses of the stable flux tubes  
 \begin{equation}
l_p \sim   \Lambda^{-1} (\Lambda L)^{- (8-2n_f)/3}  \times 
 (2 \sin \frac {p \pi}{N})^{-2},  \qquad p=1, \ldots, N-1  
\label{Eq:thickness}
\end{equation}
   We expect it to saturate to 
an $L$ independent value above the scale of gauge  symmetry restoration.  Also,  intermediate $N$-ality 
tubes seem to be much more slimmer than the small and large $N$-ality ones. 

Because of  compactification, in the weakly coupled regime, the characteristic size of the flux tubes and their tensions are no longer parametrically  related.  In the next section, we explicitly calculate  the 
string tensions. 

\subsubsection{Domain wall tensions and area law of confinement} 
\label{sec:are}
{\bf Domain walls:} Any theory which exhibits spontaneous breaking of a discrete symmetry will have discrete isolated vacua and stable domain walls which interpolate in between. QCD(adj) possesses both continuous and discrete axial chiral symmetry. As discussed in section \ref{sec:vac}, the discrete chiral symmetry is broken at any radius, thus the theory possesses stable domain walls. Note that 
in the small $S^1$ regime, the  discrete chiral symmetry $\Z_N \in \Z_{2Nn_f}  $ is intertwined with the  $\Z_N$  shift symmetry of the dual photon \ref{inter}. 

The domain wall on $\R^4$ is a three-dimensional infinite hypersurface $\R^3$. 
If    $\R^4$ is compactified  down to  $\R^3 \times S^1$ and the pattern of the discrete chiral symmetry breaking  remains invariant as a function of radius, which is the case in QCD(adj), the 
domain wall curl over itself with an $\R^2 \times S^1$ geometry.   Therefore, in the long-distance 
description, the domain wall is an  $\R^2$ filling surface embedded into $\R^3$.  Let us assume that the wall lies on  $x, y  \in \R^2$ plane and is centered at $z=0$ with a profile which  extrapolates from $z=-\infty$ to $z=+\infty$. The topological charge of such a $k$-wall (kink)  is 
 \begin{equation}
t = \int_{-\infty}^{\infty}  dz  \frac{d \sigma}{dz}   = \sigma(\infty) - \sigma(-\infty)=  \frac{2 \pi k }{N} \rho
\label{t1}
\end{equation}
 As stated earlier, the fact that the potential may be derived from a prepotential leads to the reduction of the equations of motions of the solitons to the first order (Prasad-Sommerfield type). This, combined with   Bogomol'nyi's trick, allows us to find the global minimum of the 
action in each topologically distinct sector of the effective  theory.   We have 
  \begin{eqnarray}
\langle \Omega_{k} | e^{-z H}| \Omega_0 \rangle  \equiv   \int_{\sigma_{(z= -\infty)}= 0}^ 
{\sigma_{(z= +\infty)}= \frac{2 \pi k}{N} \rho }
    D \sigma \; e^{- S(\sigma)} = e^{- {\rm Area}(\R^2)  S^{*}_k}, \qquad \Sigma\sim \R^2 \qquad 
\end{eqnarray}
Thus, the  $k$-wall  tension   is proportional  to the global minimum of the action (divided by the area of the  Area$(\R^2)$), i.e., , $T_k^{\rm DW}\equiv S^{*}_k$,,  given by  
\begin{equation}
T_k^{\rm DW}= | {\cal W} (\sigma(\infty) ) - {\cal W} (\sigma(- \infty) )|= 
| {\cal W} ( \frac{2 \pi k}{N} \rho ) - {\cal W} (0 )|
\end{equation}
in terms of prepotential. Hence, 
\begin{eqnarray}
\qquad T_k^{\rm DW}  L  =  
\frac{1}{L^2} e^{-S_0}
N | e^{i \frac{2 \pi k}{N}} - 1 | =\frac{1}{L^2}  e^{-S_0}
2 N  \sin { \frac{ \pi k}{N}}   \qquad k=1, \ldots N-1 \; . 
\end{eqnarray}
Restoring the dimensions and using the one-loop renormalization group result for the strong scale, we obtain 
\begin{eqnarray}
\qquad T_k^{\rm DW} \sim  \left(\Lambda^3(\Lambda L)^{b_0 -3} =  \Lambda^2(\Lambda L)^{2(1-n_f)/3}\right) \times  2 N  \sin { \frac{ \pi k}{N}}   \; . 
\end{eqnarray}
Note that, for $n_f=1$, this gives a new derivation of the domain  wall tension in ${\cal N}=1$ SYM, a result obtained earlier by Dvali and Shifman \cite{Dvali:1996xe}. This tension is independent of the radius. For $n_f>1$  confining gauge theories,  we expect the $L$ dependence to disappear around the strong scale, $L \Lambda \sim 1$, and expect the domain wall tension to saturate to $T_k^{\rm DW} \sim  2N  \Lambda^3   \sin { \frac{ \pi k}{N}} $ in the decompactification limit. 

{\bf Area law of confinement:}
We wish to exhibit the area law of confinement  for all but adjoint  representations ${\cal R}_p$  of the $SU(N)$ gauge group. 
 The
representations 
 of the Wilson loops  $C$ under the center group $\Z_N$ 
are in one to one correspondence with the  monodromies, 
$\int_{C'} d \sigma$ in the dual theory  \cite{Deligne:1999qp}, where  $C'$ is any closed curve 
whose linking number with $C$ is one. In QCD(adj),  
both form a  representation of  $\Z_N$.

The evaluation of   a  Wilson loop  in a representation with  charge 
$k$   under the 
 $\Z_N$ center group
   in the original theory translates into finding the field configurations for the dual scalar theory with 
monodromies  equal to $ 2\pi \mu_k$ in the dual theory where $\mu_k$ is the fundamental 
weight corresponding to external charge.    Note that  $  \mu_k = k \mu_1 + \alpha$, for some $\alpha$ valued in root lattice $\Lambda_r$, and weights differing by elements of $\Lambda_r$ are identified.  Thus, we need to find the action of 
the soliton configurations for which $\Delta \sigma= 2 \pi \mu_k $ across the Wilson loop interface, or equivalently, 
 \begin{equation}
\int_{C'} d \sigma =  \int_{z=0^-}^{z=0^+}  dz  \frac{d \sigma}{dz}   = \sigma(0^+) - \sigma(0^-)=  2 \pi \mu_k, \qquad {\rm linking }(C, C')=1
\label{t2}
\end{equation}
The reader should note that the monodromy given in \ref{t2} is not related to the topological charge of the domain wall kink given in  \ref{t1}. In particular,  
\begin{equation}
\int_{C'} d \sigma  \neq  \int_{-\infty}^{\infty}  dz  \frac{d \sigma}{dz}   
\end{equation}
although both objects, in the case of QCD(adj), are $\Z_N$ valued due to the fact 
that both the center group and discrete axial symmetry group are $\Z_N$. These two  $\Z_N$ 
are unrelated to each other.  For generic  representations, this coincidence disappears. Even 
in the case of QCD(adj), $\int_{C'} d \sigma $ is not parallel to  $\int_{-\infty}^{\infty}  dz  \frac{d \sigma}{dz} $ in the root space.  The former corresponds to interpolations  between fundamental weights on one and the same vacuum and  the latter integral   is tied with interpolations between discrete isolated vacua. 

 The expectation values of the Wilson loop fall into $N$  categories, and translate, in the path integral formulation into 
  \begin{eqnarray}
\lim_{A(\Sigma) \rightarrow \infty}  \langle W_{{\cal R}_k}(C) \rangle |_{C= \partial \Sigma} =   
 \int_{\sigma_{(z= -\infty)}= \sigma_{(z= +\infty)}}  
    D \sigma \; e^{- S(\sigma)}\Big|_{\Delta \sigma (0) = 2 \pi \mu_k} 
\end{eqnarray}
Thus, the string tension is 
  \begin{eqnarray}
T_k = \lim_{A(\Sigma) \rightarrow \infty}  
 \frac{ {\rm log} \langle W_{{\cal R}_k}(C) \rangle} {{\rm Area} (\Sigma)}   =   \min_{\sigma(z)}  
 \frac{ S(\sigma)} {{\rm Area} (\R^2)}  \Big|_{\Delta \sigma (0)= 2 \pi \mu_k}  
\end{eqnarray}
 For general $SU(N)$, we believe that the string tension in QCD(adj) should be   calculable by using the techniques  similar to  
\cite{Hollowood:1992by}.  Because of  its technical nature, we will perform this calculation in a separate publication.  The expected result is 
\begin{eqnarray}
\qquad T_k \sim  \left(\Lambda^2(\Lambda L)^{b_0 -2} =  \Lambda^2(\Lambda L)^{(5-2n_f)/3}\right) \times  2 N  \sin { \frac{ \pi k}{N}}   \; . 
\end{eqnarray}
 On the other hand, it is evident that  $T_k$ is nonzero. This is sufficient to exhibit the area law of permanent  confinement  in QCD(adj) in the $L \Lambda \ll 1$ regime, and the existence  of the linearly confining potential between two external electric sources with charges $\pm k \in \Z_N$ 
\begin{eqnarray}
V_k (R) = T_k  R,   \qquad \qquad {\rm linear \;\; confinement} 
\end{eqnarray}
We expect the tension  to saturate to a size independent  value, a c-number times $\Lambda^2$ for $L\Lambda > 1$.


 To summarize, in QCD(adj), the domain wall tensions, the string tensions,  and thicknesses   of flux tubes  (which are the inverse masses of the dual photons)    are class functions of the center group $\Z_N$. 
   The class functions depend on the 
 $N$-ality of the source,  but are  blind to the   particular representative of a class. 
  Also, exchanging (color)  source and sink is just the mirror image, and tells us that class functions must obey  $X_k= X_{N-k}$, where $X$  is any class  function.  Interesting 
  physical quantities (which are all measurable in lattice)  are the ratios of the string tensions, (inverse) string thicknesses,  and  their energy 
  densities  given by 
\begin{equation}
\frac{T_p}{T_1}= \frac{ \sin \frac {p \pi}{N}} { \sin \frac {\pi}{N}}\; , \qquad \qquad 
\frac{m_p}{m_1}= \Big(\frac{ \sin \frac {p \pi}{N}} { \sin \frac {\pi}{N}} \Big)^2,  \qquad  
 \frac{{\cal E}_p}{{\cal E}_1}= \Big(\frac{ \sin \frac {p \pi}{N}} { \sin \frac {\pi}{N}} \Big)^5 \;  . 
\label{Eq:sine} 
\end{equation}
These observable  obey 
 \begin{equation} 
 X_p \equiv X_{N+p}, \qquad X_ p= X_{N-p}, \qquad p=1, \ldots N-1
\end{equation}
Therefore, there are $\left[ \frac {N}{2} \right]$ types of flux tubes, where bracket labels the integer part of the $N/2$.  The ratio of the string tensions yields  the ``sine-law'' for the tensions.
 
In the $n_f=1$ case, the sine law  for tension 
 has previously been derived by Douglas and Shenker \cite{Douglas:1995nw}  on $\R^4$
 by deforming the $\N=2$ theory by a perturbative mass  term for the   chiral multiplet,  and  by  
 Hanany et. al. \cite{Hanany:1997hr} by realizing the same deformation in the M-theory 
 five-brane version, referred as mQCD.  \footnote{Our  result for nonsupersymmetric theories is new, and directly 
 testable on the lattice in the appropriate regime. Our derivation for the SYM is also different from earlier work \cite{Douglas:1995nw, Hanany:1997hr} and does not make any reference to supersymmetry, or the underlying theory being realizable in string theory. Due to the  generality of our approach, it is  applicable to nonsupersymmetric QCD-like theories  which are more interesting.}  
   Both \cite{Douglas:1995nw, Hanany:1997hr}   achieve  a weakly coupled $\N=1$ SYM theory 
 on $\R^4$ by adding extra matter into the theory. \footnote{An important issue here is to realize that 
 this theory is not pure $\N=1$ SYM in $\R^4$. As the authors of \cite{Douglas:1995nw} discusses,  
 this mechanism holds so long as $m/\Lambda \ll 1$,  a perturbation.  
  In order to obtain pure $\N=1$ SYM in the IR,   
 we must take $m \gg \Lambda$, which is not a perturbation, and calculational control 
 of the softly broken $\N=2$ do get lost.  Currently, there is no analytical derivation of 
 mass gap or confinement in pure $\N=1$ SYM on $\R^4$. }
 In our derivation, no  extra matter is needed.   But in order to achieve a weakly  coupled 
 formulation, we compactify the theory on  $\R^3\times S^1$ and  benefit from asymptotic freedom.  
 In both cases, the physics is rather similar,  it is spontaneously broken $U(1)^{N-1} $ gauge theory, and abelian duality  in $d=3$ and $d=4$ plays a fundamental role.   
  The formula receives $O(e^{-S_0})$ corrections, 
 which is insignificant in the $L\Lambda \ll 1$ regime, but will be essential at large radius.
  Consequently, our result does not imply that the tension will obey a sine law in large $S^1$ or in 
 $\R^4$, even in the $n_f=1$ case which is $\N=1$ pure SYM.

 {\bf Remark on other QCD-like theories:} 
Either the mass gap in the gauge sector or the area law for large Wilson loops are equally valid indicators 
of confinement for theories in which the only dynamical degrees of freedom are adjoint fermions.  For theories such as QCD with  two adjoint and one  fundamental fermions (which also breaks its gauge symmetry at small $S^1$),  the mass gap should still emerge, but area law must become  a perimeter law.   The theory should  still be confining, but the ability to form stable flux tubes must be  lost 
due to the fact that charged fermions can be pair created out of the vacuum, and break the flux tube to reduce its energy.  It would be interesting to examine this class of theories in the future. 


\subsubsection{Chiral symmetry realizations}
\label{sec:chi}
The choice of the vacuum state $|\Omega_k \rangle$    spontaneously
breaks the $\Z_N$ shift symmetry, 
which is intertwined with the $\Z_N$ discrete  chiral symmetry.  The chiral 
order parameter which is a 
singlet under continuous flavor symmetry, and which only probes the discrete chiral symmetry is 
 the determinantal condensate 
$ \det  \tr  \;  \lambda^I  \lambda^J $ in the original theory.  In the infrared of the theory on small $S^1$, 
  the  off-diagonal modes 
of the $\lambda^{I}$ are  heavy due to gauge symmetry breaking and cannot contribute to the determinantal   chiral condensate.  We may decompose  $\lambda^I= \lambda^{I,a} t^a $ into massless  components along the Cartan subalgebra and heavy off-diagonal modes,  
$\tr \lambda^I \lambda^J \sim L^{-3}
 \sum_j (\alpha_j \psi^J)  (\alpha_j \psi^J) + {\rm heavy}$, where $L^{-3}$ is due to dimensional reasons. 
The vacuum expectation value of the flavor singlet chiral condensate in 
$SU(N)$ QCD(adj) with $1 \leq n_f \leq 4$ flavor can be found 
by integrating over the zero-mode wave functions (which are essentially proportional to monopole profiles) in the background of a monopole in the small $S^1$ regime, where the gauge symmetry 
is broken.  On large $S^1$, we do not know a reliable  analytical technique in the $1 <n_f \leq 4$ case 
to evaluate the condensate.  However, we expect  the modulus of the chiral condensate to saturate to a c-number times  $\Lambda^{3n_f}$.  Consequently, 
\begin{equation}
\langle \Omega_k | \det  \tr  \;  \lambda^I  \lambda^J |\Omega_k \rangle   
\sim \left\{  \begin{array}{ll} 
 \Lambda^{3n_f} 
 (\Lambda L)^{ \frac{11}{3}(1-n_f)}  e^{\frac{i 2 \pi k}{N}}  &  \qquad L \ll L_c  \cr 
  \Lambda^{3n_f}   e^{\frac{i 2 \pi k}{N}} ,  &  \qquad    L> L_c  
  \end{array}
  \right.
\label{Eq:detconden1}
\end{equation} 
where  the  phase is $\Z_N$ valued.  In the $n_f=1$ case,  this produces the correct $L$  independence 
of chiral condensate (which is due to supersymmetry) \cite{Davies:2000nw}, and $N$  isolated vacua.  
We believe  that  the scale at which the determinantal condensate becomes $L$ independent is the scale of the gauge symmetry restoration.

\begin{figure}[t]
{
  \parbox[c]{\textwidth}
  {
  \begin{center}
  \includegraphics[width=2in]{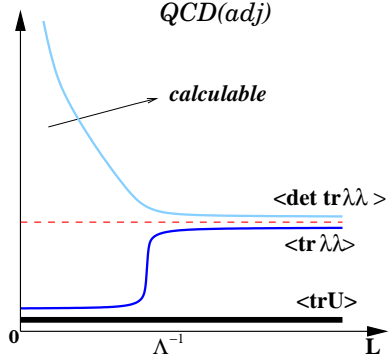}
  \caption
    {%
The cartoon of the  behavior of the center, discrete and continuous chiral symmetry realization in QCD(adj), for $SU(N)$ where $N=$ few, $n_f=2$ and $n_f=1$ ($\N=1$ SYM).  
 The spatial  center symmetry is unbroken at any $L$ in both cases $\langle \tr U \rangle =0$. 
 In $n_f=2$, 
the continuous chiral symmetry is unbroken at small $S^1$ and broken at large $S^1$, and discrete chiral symmetry is always broken. 
The red (dotted) line is the chiral condensate in $\N=1$ SYM, and the  discrete chiral symmetry is 
always broken.  
In the small $S^1$ regime, the string tensions and thicknesses (the inverse mass gap in gauge sector) are calculable, and $n_f=2$ theory exhibits confinement without continuous chiral symmetry breaking. 
The lines slightly on top of the  horizontal  axis are all zero and  are split to guide the eye.     
   \label{fig:tension1} }
  \end{center}
  }
}
\end{figure}

In the far infrared of the QCD(adj), since $\sigma$ is massive, the long-distance theory further reduce 
to a purely fermionic theory, which schematically looks like an NJL-type Lagrangian:
 \begin{equation}
  L_{\rm NJL}=  \sum_{j=1}^{N}
\left[  i\bar \psi^I_{j} \gamma_{\mu} \partial_{\mu} \psi^I_j   + c  e^{-S_0}  
 ( \det_{I, J} \alpha_j \psi^I  \alpha_j \psi^J +  \rm c.c.)  \right]
\label{Eq:NJL}
\end{equation}
The Lagrangian is invariant under  $SU(n_f) \times \Z_{2n_f} $ chiral symmetry.  The $\Z_{2n_f}$  is the unbroken subgroup of the  $\Z_{2Nn_f}$ discrete symmetry. We wish to know whether the continuous chiral symmetry is broken spontaneously.  

At small $S^1$,  we believe the continuous chiral symmetry is unbroken, based on studies 
on related  $d=3$ dimensional NJL-type  models. Such models   have generically   a weakly coupled 
chirally symmetric phase and a chirally asymmetric strong coupling phase. 
(See the review in Ref.\cite{Rosenstein:1990nm}).
Our  dimensionless coupling constant is $g \sim e^{-S_0}$,   far too small to induce a chiral 
transition.  
Hence, the chiral symmetry must  be unbroken, and there must be massless fermions (protected by chiral symmetry) in the spectrum 
 within the region of validity of our long-distance effective theory,  ($L \Lambda \ll 1$). 
We believe the naive extrapolation of the NJL Lagrangian 
Eq.\ref{Eq:NJL} will  exhibit   the continuous chiral transition in an 
expected regime of the underlying  QCD theory.   (See fig.\ref{fig:tension1}.)
 However, this will happen outside the region of validity of our effective theory.  
Consequently, this does not tell us that the monopole operator  is the sole origin of  the  continuous chiral symmetry breaking, even though it is the origin of the discrete chiral symmetry breaking in the small $S^1$ regime.  In the large $S^1$ regime,   non-dilute monopoles  
with fermionic zero-modes  play the major role in continuous chiral symmetry breaking. 

The absence of the continuous chiral symmetry breaking in weak coupling regime 
 can also be seen by an independent argument.  In the small $S^1$ regime where theory is weakly coupled, we have control over all nonperturbative objects.  A BPS or KK monopole, which may in principle contribute to the condensate, has a minimum of $2n_f$ fermionic zero-modes.  
However,  our order parameter     $ \tr  \;  \lambda^I  \lambda^J $ can only soak up two zero-modes. 
This implies it cannot acquire a non-trivial vacuum expectation value. 
The minimal operator which may acquire a condensate must have $2n_f$  fermion insertion, and 
this is indeed the determinantal condensate $ \langle \det  \tr  \;  \lambda^I  \lambda^J \rangle $. 
The reliability of this argument is tied with weak coupling, and in fact, it does not hold at strong coupling.

 At large $S^1$ (and $\R^4$), the common lore is that 
the chiral symmetry is spontaneously broken down to $SO(n_f) \times \Z_2$ by the formation of the 
chiral condensate 
\begin{equation}
\langle \Omega_k |  \tr  \;  \lambda^I  \lambda^J |\Omega_k \rangle   
\sim \left\{  \begin{array}{ll} 
 0 &  \qquad L < L_c  \cr 
  \Lambda^{3}  e^{\frac{i 2 \pi \kappa}{Nn_f}} ,  &  \qquad    L> L_c  
  \end{array}
  \right.
\label{Eq:conden}
\end{equation} 
Consequently,  there must exist  $N$  isolated coset spaces each of which is  $SU(n_f)/SO(n_f)$.   
 In this expression,  $\kappa$ ranges in   $[0,  N n_f )$.  
      Denote $\kappa= \eta N  + k$ 
 where $k=0, \ldots N-1$   and $\eta=0, \ldots n_f-1$.  For a given $k$, there are $n_f$ many $\eta$ 
 for which the determinant of the condensate is invariant.  Thus, they reside in the same coset space, and  there are consequently $N$ isolated coset spaces.

 The continuous  chiral transition in QCD(adj) is very different from its thermal counterparts.  In particular, 
 it  occurs  in the absence of any change in its spatial center symmetry realization.  
This is a quantum phase transition at absolute 
zero temperature, induced solely due to quantum fluctuations. We do not know the order of the phase transition. 

Finally, we wish to conjecture  that the scale of the chiral phase transition  $L_c$  in QCD(adj) is associated  with the restoration of the spontaneously broken gauge symmetry. Consequently, 
we believe that the chiral symmetry breaking is a strong coupling phenomena.  Confinement is not necessarily so\footnote{In subsequent work, I showed the natural scale of chiral symmetry breaking at arbitrary $N$ is $\Lambda^{-1}/N$. Figure \ref{fig:tension1} is for $N=$few for which there is no parametric separation between $\Lambda^{-1}$ and $\Lambda^{-1}/N$.}. 

\subsection{Noncompact versus compact adjoint Higgs, final pass}
\label{sec:noncom}
Let us reconsider the $SU(N)$ gauge theory with a  {\it  noncompact}   adjoint   Higgs field and with 
one Dirac  fermion in adjoint representation on $\R^3$. (Multiflavor generalization is obvious.) 
  The theory possess a $U(1)$ fermion number  symmetry.  
 The generalization of the argument of Ref.\cite{Affleck:1982as} shows that the $U(1)$ symmetry is spontaneously broken, and consequently, there only exists one gapless excitation by Goldstone's 
 theorem.   The other  $N-2$ photons of the spontaneously broken gauge symmetry must acquire 
 masses. We wish to know how  this is  realized in the microscopic description.

When the $SU(N)$ gauge symmetry breaks down to $U(1)^{N-1}$ via an noncompact adjoint Higgs field rather than a compact one (which was the case in QCD(adj)), 
monopoles only come in $N-1$ varieties.  The 
KK monopole is now absent. 
We may still define the magnetic   bions in the spontaneously broken 
$SU(N)$ gauge theory for $N\geq 3$, but there are only $N-2$ of them.   As before, a
bion is a  bound state of the monopole associated with magnetic charge $\alpha_i$ and anti-monopole associated  with charge $- \alpha_{i+1}$ with null topological charge.  
 The magnetic 
charge of a bion is 
 \begin{eqnarray}
 Q_i= \alpha_i - \alpha_{i-1}, \qquad   i=2, \ldots N-1
\end{eqnarray}
Hence, there are only $N-2$ types of magnetic bions. In other words, the absence of the 
$\alpha_N \equiv 
\alpha_0$ KK monopole removes two would-be bions of the compact theory.  
 Thus, the potential for the  
$\sigma $ field is a sum over $N-2$ bions and their conjugates given by 
 \begin{eqnarray}
V(\sigma)=  - e^{-2S_0} \sum_{i=2}^{N-1}   \left( 
 e^{i Q_i \sigma }    + c.c
 \right)   
 \end{eqnarray}
 The potential   generates mass terms only for $N-2$ dual photons. The massless 
 photon is the Goldstone boson.  
 Equivalently, we may say the sum in the prepotential is restricted to the root system $\Delta_0$,  
$ {\cal W}(\sigma)= e^{-S_0} \sum_{ \alpha_{i} \in \Delta_{\rm }^{0}  } e^{  i \alpha_i \sigma } \; $, and 
from the study of the analogous supersymmetric theory, we know that the reduction from affine Toda to 
nonaffine Toda renders the gapped  theory gapless \cite{Davies:2000nw, Katz:1996th}.  
 
\section{Outlook: Confinement and non-self-dual topological excitations}  
 \label{sec:con}
A  microscopic derivation of the mechanism which provides confinement in QCD(adj) quantized on  small $S^1 \times    \R^3$ is given.   This is a QCD-like theory with no elementary 
scalars in its Lagrangian, and no special properties such as supersymmetry (except the $n_f=1$ case).
We believe the solution  provides a significant   contribution to our current understanding of QCD-like gauge theories, and teaches us many valuable lessons.    We also found    the   
underlying dynamical  reasons behind the lack of confinement
in  Yang-Mills noncompact Higgs systems  with adjoint fermions formulated on  $\R^3$.  
   Let us quote our main  result for the  $SU(2)$  gauge group:

\begin{itemize} 

\item{New  non-self-dual topological excitations that we referred to as 
magnetic bions  exist in the QCD(adj) vacuum  
and are the source of confinement. A mechanism by non-self-dual excitations was not suspected  
in QCD-like theories by the wisdom gained from other analytically solvable theories, such as Polyakov model or Seiberg-Witten theory. Even the existence of such stable topological excitations is surprising as they are topologically neutral, just like perturbative vacuum! But they carry a magnetic charge. }

\item{QCD(adj) exhibits permanent confinement even at arbitrarily weak  coupling (small $S^1$).   In other words,  in asymptotically free confining gauge theories, confinement is not  {\it necessarily} a strong  coupling phenomena. }

\item{
 In the presence of massless adjoint dynamical fermions,   the monopole operators 
  must have  compulsory  fermion zero-mode attached to them.  Hence, they induce 
 fermion-fermion  and fermion-dual photon interactions, neither of which   can appear in the bosonic potential of the dual photon.  Our arguments  rule out monopoles and monopole condensation as the microscopic mechanism  of the  confinement in  QCD-like theories with {\it dynamical fermions} in general.   
 }

\item{
The  beautiful and qualitative idea of dual superconductivity is quantitatively realized in the vacuum of 
QCD(adj), but not in terms of self-dual monopoles, or instantons.  
Non-self-dual  magnetic bions with magnetic and topological charge  $(\pm 2, 0) $ generate a mass gap in  the gauge sector and confinement. }

\item{ Magnetic bions are composites of BPS and $\overline{\rm KK}$ monopoles, and their  stability is due to a dynamical fermionic   pairing mechanism.  
The repulsive Coulomb repulsion between the bion constituents  [with charges $(1, +\half)$ and 
$(1, -\half)$ ]  is overwhelmed by a  attractive  logarithmic force.   
The pairing mechanism responsible for the bound state is  induced by 
  $2n_f$-fermion exchange in $n_f$ flavor theory.}


\item{ This rationale also explains why the Yang-Mills with {\it noncompact}
 adjoint Higgs field and  adjoint 
fermions does not confine on $\R^3$  despite the presence of monopoles. The same rationale is also true 
for $\N=2$ SYM on $\R^3$.  These are examples as important as QCD(adj) itself, because we believe 
it is equally important to understand the lack of confinement in order to understand  confinement. 
 }

\item{In the general $SU(N)$ case, we demonstrated the area law of confinement for Wilson loops in arbitrary representations.  The dual theory hints at  an integrable (generalized Toda) system behind QCD(adj), in the $e^{-S_0}$ expansion of the action at order $e^{-2S_0}$.     We do not know whether this  extends to higher order 
if we were to find higher-order terms in $e^{-S_0} $ expansion.   We also do not know whether 
there may be integrability behind QCD(adj) on   $\R^4$.} 
\end{itemize} 

We wish to express that we are  optimistic of future  progress which will reveal more on the inner goings-on of general QCD-like theories: 

{\it Incorporating fundamental representation fermions:}   For example, in a theory with  two adjoint and 
one fundamental fermions ({\bf mixed action}),  the back-reaction of the fundamental 
fermion is insufficient to induce center 
symmetry breaking in the small $S^1$ regime. This theory has 
  both magnetic monopoles and  massless electric charges within the weak coupling  regime examined in this paper. This system should teach  us something which may be  relevant  to  the 
  real QCD. 
  Unfortunately, our techniques are not directly applicable to  pure Yang-Mills or QCD with fundamental fermions due to breaking of (temporal or  spatial)
   center symmetry at small $S^1$. 

{\it Confinement on QCD-like theories on $\R^4$:} The techniques of this paper are strictly 
valid in the gauge symmetry broken phase of the QCD(adj). However, we believe that certain assertions  
are generalizable to $\R^4$, and direct  progress will occur  in  QCD(adj) on $\R^4$,  where strong coupling necessarily occurs. 

{\it Lattice gauge theory:} Many assertions made in this paper are directly testable in lattice simulations 
with available technologies.  In particular, the string tensions  and characteristic sizes of flux tubes \ref{Eq:Spec}, \ref{Eq:sine} can be extracted from the lattice simulations of QCD(adj)  as in 
\cite{Lucini:2004my}.   
QCD(adj) also undergoes a zero temperature quantum chiral transition in the absence of any change in center symmetry realization. This should be directly testable on  the lattice by modifying the existing simulations (such as  \cite{Karsch:1998qj}) appropriately.  
It would also be  useful to construct the duality between  QCD(adj) on $\R^3 \times S^1$ with Lagrangian \ref{eq:Lagrangian} and  dual QCD defined in   \ref{Eq:dQCD2}  directly in lattice formulations.

\acknowledgments
I am grateful to Misha Shifman and Du{\v s}an Simi{\'c} for enlightening discussions. I thank 
 Ofer Aharony,  Tom DeGrand,  Tim Hollowood, Shamit Kachru, Michael Peskin, Erich Poppitz, 
   Steve Shenker, David Tong,  Larry Yaffe, and my colleagues 
 at SLAC, Stanford and Boston University for communications  related to this work. In the second version of this work, I benefitted from a  conversation with Erich Poppitz. I would like to thank him for explaining to me the non-self-dual nature of the magnetic bions. 
 This work was supported by the
U.S.\ Department of Energy Grant DE-AC02-76SF00515

{\bf Note added:} There was  a large time delay between the arXiv version of this paper and its submission to a journal. In the meantime, new useful techniques, such as center stabilizing double-trace deformations,   which allows a smooth connection of small and large $S^1$ physics, 
and the  relevant index theorem for generic topological excitations on  $S^1 \times \R^3$
have been found. These techniques  enabled us to study non-perturbative dynamics of all vector-like and even chiral theories on $S^1 \times \R^3$.  In all  chiral theories and QCD-like theories 
with  two index matter representations, we now understand that magnetic bions or  similar composite non-self-dual excitations are the root cause of confinement.  For a review of these developments and related works, see the recent preprint \cite{Poppitz:2009uq}.

\bibliographystyle{JHEP} 

\bibliography{confinement1}

\end{document}